\documentclass[11pt]{article}

\usepackage[letterpaper]{geometry}  
\usepackage[english]{babel}         
\usepackage{graphicx}               
\usepackage{float}  
\usepackage{subcaption}
\usepackage[colorlinks=true, allcolors=blue]{hyperref}  
\usepackage{amsmath}
\usepackage{mathrsfs}
\usepackage{amsfonts}
\usepackage{csvsimple}
\usepackage{tabularx}
\usepackage{amsthm}
\usepackage{authblk}
\usepackage{tikz}
\usepackage[dvipsnames]{xcolor}
\usetikzlibrary{arrows.meta, positioning}

\numberwithin{equation}{section}
\graphicspath{ {images/} } 

\begin{document}

\title{\bf\Large {Quasi-exactly solvable deformations of quantum systems associated with exceptional orthogonal polynomials}}

\author[1]{Siyu Li \footnote{Siyu.Li@latrobe.edu.au}}
\author[2]{Ian Marquette \footnote{i.marquette@latrobe.edu.au}}
\author[3]{Sarah Post \footnote{spost@hawaii.edu}}
\author[4]{Yao-Zhong Zhang \footnote{yzz@maths.uq.edu.au}}

\affil[1]{\em Department of Mathematical and Physical Sciences, La Trobe University, Bundoora, VIC 3086, Australia}
\affil[2]{\em Department of Mathematical and Physical Sciences, La Trobe University, Bendigo, VIC 3552, Australia}
\affil[3]{\em Department of Mathematics, University of Hawaii, Honolulu, HI 96822, United States of America}
\affil[4]{\em School of Mathematics and Physics, The University of Queensland, Brisbane, QLD 4072, Australia}


\maketitle

\begin{abstract}

\noindent Quasi-exactly solvable (QES) deformations of harmonic and singular oscillators have been widely studied via a range of approaches. In this paper, we obtain families of new QES deformations of solvable quantum systems associated with exceptional orthogonal polynomials (EOPs). The construction builds upon the theory of Darboux-Crum transformations and the classification of exactly solvable quantum systems associated with EOPs of Hermite type. It is shown that the deformations break the exact solvability of the undeformed systems and introduce model parameters into the deformed systems that permit the existence of a finite number of polynomial solutions whose roots satisfy systems of algebraic equations. The new families presented and studied consist of deformations of quantum systems related to Hermite EOPs of type III with arbitrary codimensions. We present polynomial and rational deformations and analyze, in each case, the conditions for quasi-exact solvability in terms of Bethe ansatz equations and parameter constraints. In general, the structures of the underlying polynomial solutions are no longer associated with well-known classical orthogonal polynomials. So for the general cases, we mainly focus on presenting the new approach as well as the Bethe ansatz equations and the constraints for model parameters. As applications, we construct particular families of QES deformations related to solvable models allowing one, two, and up to four gaps, and   obtain the closed form expressions for their wavefunctions and spectra. 
We analyze the existence of QES solutions in the spaces of model parameters, providing information on the number of solutions for given parameters. 

\end{abstract}

\section{Introduction}
Among exactly solvable (ES) quantum systems, the harmonic oscillator and its algebraic structure are undoubtedly the most studied. Despite their simplicity, they admit many interesting solvable  deformations. Examples of such deformations include rational deformations via Wronskians of Hermite/pseudo Hermite polynomials, nonlinear deformations via Painlev\'e, anharmonic deformations, and $q$-deformations \cite{bender1969anharmonic, banerjee1978general, bonatsos1992classical, lorek1997q, sogami2003q, carinena2017rational}. In many of these deformed systems, the existence of ladder operators can be established  \cite{carinena2017abc,marquette2013newladder,bosso2018generalized} and used to provide their closed-form solutions and spectra  \cite{dong2005exact, aouda2020ladder}. In particular, state-adding deformations of harmonic oscillators via the Darboux-Crum transformation play an important role in the construction of novel families of exact solvable potentials \cite{cooper1995supersymmetry, david2010supersymmetric, sasaki2014exactly}. 

Exceptional orthogonal polynomials (EOPs) were originally introduced within the framework of Sturm-Liouville eigenproblems \cite{gomez2007quasi,gomez2009extended}. They naturally appear in the study of rational extensions of exactly solvable quantum systems.  The essential property of EOPs is that they form a complete set of orthogonal polynomials and admit missing degrees. They generalize classical families such as Hermite, Jacobi and Laguerre polynomials. Similarly to classical orthogonal polynomials, the exceptional families are also connected through nontrivial interrelations \cite{quesne2024connecting}. They are related to the confluent limits of generalized Sch\"ur polynomials \cite{grandati2014exceptional} and can be extended to all partitions via complex integration contours and non-positive Hermitian products \cite{haese2016complex}. Moreover, it is known that an EOP can be obtained from a classical system via a Darboux-Crum transformation, also know supersymmetry transformation (SUSY) in quantum mechanics \cite{gomez2010exceptional, gomez2013conjecture, garcia2019bochner}.  From the perspective of supersymmetric quantum mechanics, Darboux-Crum transformations provide a systematic way to generate new quantum systems from classical exactly solvable models, including those related to EOPs \cite{quesne2008exceptional, quesne2009solvable, sasaki2010exceptional, quesne2011higher, marquette2013, marquette2013new, marquette2014, agboola2014, gomez2013rational, quesne2004more}. Those models display properties that make them interesting from point of view of applications, such as complex patterns of irreducible representations, spectrum with layer structures and non uniqueness of the recurrence relations and ladder operators. This last property is unique to systems associated with EOPs. Classification of non equivalent ladders has not been addressed in general. 

Recently, deformations that partly break the exact solvablity property of a solvable model have attracted much attention. Such deformations result in systems which are partly or quasi-exactly solvable (QES). By QES, one refers to a model in which only part of its spectrum and corresponding wave functions can be obtained analytically and algebraically \cite{turbiner1988quasi, turbiner2016one, ushveridze2017quasi}. Unlike the ES models, parameters in a QES model satisfy certain constraints which allow its Hamiltonian to preserve a finite number of polynomial spaces called flags. Polynomial solutions of QES models are not expressible in terms of classical orthogonal polynomials \cite{finkel1996quasi, bender1996quasi} and  can be obtained using the functional Bethe ansatz method  \cite{zhang2012}. The closed-form eigenfunctions and spectrum of anharmonic oscillator with polynomial type potentials have been studied in, e.g. \cite{agboola2013novel, agboola2013exact} \cite{klink2023polynomial, schweiger2025polynomial}. 

In this paper, we develop a framework for constructing new QES deformations of exactly solvable systems. The key aspect of our approach is to find QES deformations of the supersymmetric partners of ES systems. This is achieved by first applying Darboux-Crum type transformations to ES systems and then deforming the transformed systems. So, this approach  is different from the one in our recent work \cite{li2025new} in which new QES systems are obtained via applying supersymmetry directly to known QES systems. We obtain polynomial and rational deformations of ES Darboux/multi-step transformed systems connected with multi-indexed EOPs Hermite of type III. We demonstrate how the approach allows us to prove quasi-exact solvability properties and derive closed-form solutions and spectrum using a combination of Bethe ansatz and recurrence relations. The framework provides new families of QES potentials and corresponding analytic solutions.

The paper is organized as follows. In Section 2, we recall the multi-step transformed harmonic oscillator and the anharmonic deformation of the harmonic oscillator. The first is exactly solvable deformation, and the second is the well-known QES deformation of the harmonic oscillator. We then illustrate our new approach and present new families of QES deformations. We show that how the additional terms in the potentials and their compatibility with quasi-exact solvability are obtained through parameter constraints and a set of algebraic equations, i.e. the so-called Bethe ansatz equations. This is done for both polynomial and rational deformations. In Sections 3, we give applications of our method to specific examples of QES deformations. The allowed values for the model parameters can be determined in the parameter space using numerical techniques. Our approach allows us to get insight into how the number of solutions changes with the change of deformation constraints.  We conclude the paper in section 4.

\section{New families of QES potentials}
The main aim of this section is to develop a new approach for the construction of QES systems. The framework is based on the state-adding approach and anharmonic deformations of ES systems. We begin by recalling some known formulas from the Wronskian formulation of state-adding transformations and multi-step oscillators, to be used as input.  The resulting QES deformation families will not retain the underlying SUSYQM transformation.

\subsection{New method for constructing QES potentials}
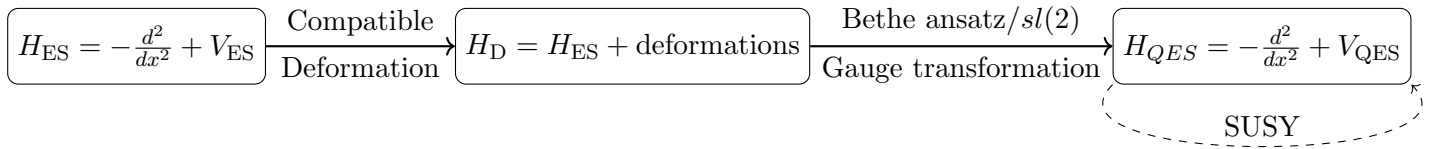
\begin{figure}[htbp]
\centering
\begin{tikzpicture}[
block/.style={draw, rectangle, rounded corners, minimum width=3cm, minimum height=1cm},
arrow/.style={->, thick}
]

\node[block] (A) {$H_{\text{ES}} = -\frac{d^2}{dx^2}+V_{\text{ES}}$};

\node[block, right=2.5cm of A] (B) {$H_{\text{D}}=H_{\text{ES}}+\text{deformations}$};

\node[block, right=4cm of B] (C) {$H_{QES}=-\frac{d^2}{dx^2}+V_{\text{QES}}$};

\draw[arrow] (A) -- node[above]{Compatible} node[below]{Deformation}(B);

\draw[arrow] (B) -- node[above]{Bethe ansatz/$sl(2)$} node[below]{Gauge transformation}(C);

\draw[->, dashed]
(C.south west) to[out=-135, in=-45, looseness=1]
node[above] {SUSY}
(C.south east);

\end{tikzpicture}
\caption{Construction of the QES systems. }
\label{fig:QES-diagram}

\end{figure}

The construction of QES potentials has been studied widely by different direct methods. The use of supersymmetric transformation is more recent. It is applied as illustrated in Fig.\ref{fig:QES-diagram}. In the diagram, the first block represents an initial Hamiltonian with the exact solvability property, the second its superpartner and the third a deformation which in the case of the harmonic oscillator will be anharmonic. In this setting, QES deformation terms are added to an initial ES system. After selecting a compatible deformation,  a set of constraints on model parameters and Bethe ansatz equations will determine the polynomial solutions of the deformed systems (after appropriate gauge transformations). In general, only some of these systems would have a hidden algebra symmetry.  This is illustrated by the second arrow in the diagram. Very recently  \cite{li2025new}, the problem of applying supersymmetric transformation state deleting (i.e. Krein-Adler ) was studied systematically for QES systems with a $sl(2)$ hidden symmetry. Note that the approach is also applicable to any QES systems with a finite number of states. The dashed line indicates the possibility of constructing new QES systems using such an approach.

\vspace{10pt}

\vspace{10pt}

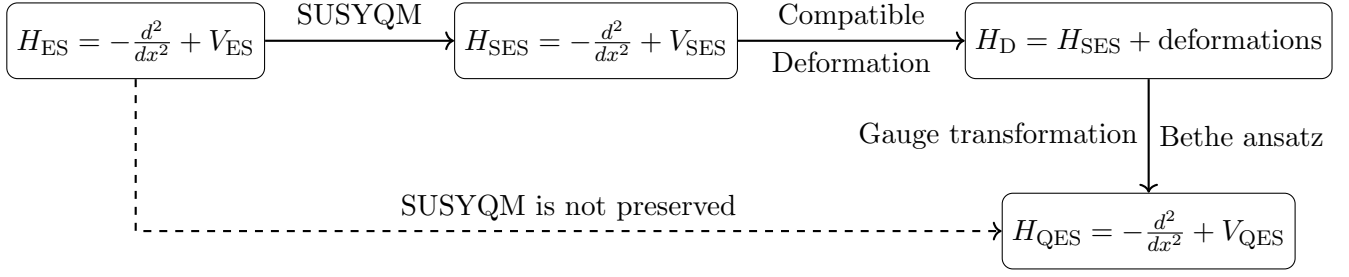
\begin{figure}[htbp]
\centering
\begin{tikzpicture}[
block/.style={draw, rectangle, rounded corners, minimum width=3cm, minimum height=1cm},
arrow/.style={->, thick}
]

\node[block] (A) {$H_{\text{ES}} = -\frac{d^2}{dx^2}+V_{\text{ES}}$};

\node[block, right=2.5cm of A] (B) {$H_{\text{SES}}=-\frac{d^2}{dx^2}+V_{\text{SES}}$};

\node[block, right=3cm of B] (C) {$H_{\text{D}}=H_{\text{SES}}+\text{deformations}$};

\node[block, below=1.5cm of C] (D) {$H_{\text{QES}}=-\frac{d^2}{dx^2}+V_{\text{QES}}$};

\draw[arrow] (A) -- node[above]{SUSYQM} (B);

\draw[arrow] (B) -- node[above]{Compatible} node[below]{Deformation}(C);

\draw[arrow] (C) -- node[right]{Bethe ansatz} node[left]{Gauge transformation}(D);

\draw[arrow, dashed] (A) |- node[pos=0.75, above]{SUSYQM is not preserved}
(D);

\end{tikzpicture}
\caption{New approach of the SUSYQM construction for QES systems. }
\label{fig:SUSY-QES-diagram}

\end{figure}

This formalism, i.e. of applying SUSYQM to known QES systems, is difficult to implement in general in a systematic way, in particular for higher order SUSYQM, because of the complicated structures of the states and their zeros of the initial QES systems. The approach taken in this paper for establishing new families of QES potentials changes the order of operations,  as shown in Fig.\ref{fig:SUSY-QES-diagram} below which illustrates the key steps of the approach. As seen from the diagram, we first carry out the SUSY transformation to the starting system, which can be done in a systematic way, then do compatible deformations and the implement the Bethe ansatz method. 
 In such a setting, the SUSYQM transformation for the initial system is not preserved and thus the deformed system is not a superpartner of the initial Hamiltonian.  We believe that this approach potentially has wide applicability, and allow us to derive several new families of QES systems which may have applications in different contexts. 

\vspace{10pt}

\vspace{10pt}


\newpage
Let us review some of the particulars of the steps in the above approach. We first establish the Hamiltonian $H_{\text{SES}}$ via SUSYQM, using a subset of solutions for an ES system, including physical and non physical solutions. These are referred to as ``seed solutions" and denoted $\phi_i$. The intertwining between $H_{\text{ES}}$ and $H_{\text{SES}}$ is (\ref{intertwine supercharge}). 
\begin{equation}\label{intertwine supercharge}
 A H_{ES}  = H_{SES} A,\quad A= A_NA_{N-1}.... A_2A_1,   
\end{equation}
where $A$ is $n$th order supercharge. $A_i$ is constructed by the superpotential $W_i$ determined by the Wronskian of the seed solutions. 
\begin{equation}\label{superpotential partner wavefunction}
 A_i= \frac{d}{dx}  + W_i ,\quad W_i= -\frac{d}{dx} \log{\Phi_i} ,\quad \Phi_1=\phi_1,\quad \Phi_i=\frac{{\cal W}(\phi_1,...,\phi_i)}{{\cal W}(\phi_1,...,\phi_{i-1})},\quad i=2,...,N.    
\end{equation}
 In this framework, which is well established, a chain of ES superpartners is created. Here we are only concerned with the final superpartner $H_{\text{SES}}$ that has  certain regularity property. This depends on the initial systems, singularities, and boundary conditions. This framework ensures that $H_{\text{SES}}$ also has the exact solvability and may have an infinite set of square integrable states (with at most $n$ states added in the state adding approach or $n$ states deleted in the state deleting approach).

We will use $H_{\text{SES}}$ to construct $H_{\text{D}}$ by adding suitable QES deformations. The types of deformations are those that would be compatible with the initial Hamiltonian $H_{\text{ES}}$ and our criteria will be that $H_{\text{QES}}$ will contain free model parameters which satisfy suitable constraints and its solutions can be determined by the Bethe ansatz equations. Obviously, the possible deformations will depend on initial ES systems and the types of functions (e.g. rational or trigonometric functions) in the potentials. In this paper, we will only focus on polynomial and rational type deformations.

In order to apply the Bethe ansatz method, one needs to bring $H_{\text{D}}$ into a standard form; this is achieved by using appropriate gauge transformation.  The Bethe ansatz method determines the constraints of model parameters and give a set of algebraic equations for the Bethe roots. The parameters in the potentials of $H_{\text{D}}$ that satisfy the constraints make the model quasi-exact solvable. Solving Bethe ansatz equations is known to be difficult, in particular as we seek higher order excited states, even with numerical approaches. The parameters allowed for $H_{\text{QES}}$ are provided in the parameter space using different numerical techniques. We will focus on determining the existence of closed-form polynomial solutions for a broad range of parameters. The analytical wave functions of $H_{\text{QES}}$ consist of the gauge factors and the closed-form polynomials determined by the Bethe ansatz equations. 


\subsection{SUSYQM Example: Generalized $n$th order state-adding of harmonic oscillator}
This section describes an example of the first arrow in Fig.\ref{fig:SUSY-QES-diagram}. 
We start from the $n$th-order state-adding transformation of the harmonic oscillator. The wavefunction of the harmonic oscillator can be written as \cite{marquette2013,marquette2014}
\begin{equation}\label{SUSY ES harmonic original wavefunctions}
    \psi_n(r) = H_n(r) e^{-\frac{1}{2}r^2},
\end{equation}
where $H_n(r)$ is the $n$th-degree Hermite polynomial. For the state-adding SUSY transformation, we use the nonphysical state as the seed solution. Changing the variable $r$ to $ir$, the nonphysical state $\phi(r)$, which is also the eigenstate of the Hamiltonian, can be written as
\begin{equation}\label{SUSY ES harmonic original nonphysical states}
    \phi_m(r)={\cal H}_m(r)e^{\frac{1}{2}r^2},\quad {\cal H}_m(r)=(-i)^m H_m(ir), \quad m=2,4,6,\cdots
\end{equation}
where ${\cal H}_m(r)$ is a pseudo-Hermite polynomial of $m$ degree. For the first order SUSY transformation, an odd $m$ gives a seed with zeros and is not admitted as a nodeless seed solution. In the higher order SUSY transformation, the odd $m_i$ are admissible to provide nodeless seed solutions when the value-selection rule below is satisfied. 
\begin{equation}\label{SUSY ES harmonic original pseudo-Hermite polynomial}
    {\cal H}_m(r)=(-i)^m H_m(ir)=m!\sum_{p=0}^{[m/2]}\frac{(2r)^{m-2p}}{p!(m-2p)!},\quad m=0,1,2,\cdots
\end{equation}
The seed functions $\phi_m(r)$ are relabeled in the $N$th order SUSY transformation to indicate the degrees of the seed solutions, i.e.
\begin{equation}
(\phi_1,\phi_2,\cdots,\phi_N)\rightarrow (\phi_{m_1},\phi_{m_2},\cdots,\phi_{m_N}),
\end{equation}
where $m_1<m_2<\cdots<m_N$. The $N$th order partner potential $V^{(N)}(r)$ is built by the Wronskian of $N$ seed solutions.
\begin{equation}\label{SUSY ES harmonic nth SUSY partner potential}
    V^{(N)}(r)=r^2-2\frac{d^2}{dr^2}\ln {\cal W}(\phi_{m_1},\phi_{m_2},\cdots,\phi_{m_N})=r^2-2N-2\frac{d^2}{dr^2}\ln{\cal W}({\cal H}_{m_1},{\cal H}_{m_2},\cdots,{\cal H}_{m_N}), 
\end{equation}
where ${\cal W}(\phi_{m_1},\phi_{m_2},\cdots,\phi_{m_N})$ and ${\cal W}({\cal H}_{m_1},{\cal H}_{m_2},\cdots,{\cal H}_{m_N})$ are the Wronskians of the seed functions $\phi_{m_i}$ and the pseudo-Hermite polynomials ${\cal H}_{m_i}$, respectively. The $-2N$ can be absorbed into the energy and we use the corresponding shifted energy convention in the later deformed potentials.
To avoid new singularities appearing in $V^{(N)}(r)$, the degree $m_i$ in the Wronskian should satisfy the following value-selection rule: when the degree of the pseudo-Hermite polynomial $m_i$ is even (odd) integer, the subscript $i$ of $m_i$ will be odd (even). Therefore, ${\cal W}_N={\cal W}({\cal H}_{m_1},{\cal H}_{m_2},\cdots,{\cal H}_{m_N})$ has even parity and the degree of the ${\cal W}({\cal H}_{m_1},{\cal H}_{m_2},\cdots,{\cal H}_{m_N})$ is always even. 
\[{\cal W}_N(r)={\cal W}_N(-r),\qquad \text{deg}({\cal W}_N)=\sum_{i=1}^N m_i-\frac{N(N-1)}{2}.\]

Different types of higher-order SUSY transformations can be written using a Wronskian formulation for the superpartner and its eigenstates \cite{BERMUDEZ2012692,BERMUDEZ201635}. This includes SUSY transformations such as Krein-Adler and Darboux-Crum transformations \cite{Odake2013}. The distinction between the different types of SUSY transformations has its origin in the type of seed solutions, i.e. the type of eigenstates involved in the Wronskian formula, for example, physical or non-physical states, with energies below or above the ground-state energy of the initial Hamiltonian. The Wronskians and various determinantal identities of the classical orthogonal polynomials can then be exploited to gain further insight into the structure of the wavefunctions and their connection to exceptional orthogonal polynomials \cite{Gomez10.1063}. These SUSY transformations allow one to obtain new families of exactly solvable rational deformations. There are various properties underlying these exactly solvable extensions that are of interest, such as the characterisation of the zeros of the related EOPs \cite{Ho2012,KUIJLAARS201528}. For deformations of the harmonic oscillator, the zeros of the EOPs are related to Young diagrams \cite{FELDER20122131}. This connects with the present work, as we intend to obtain deformations of various exactly solvable systems into quasi-exactly solvable systems, and their zeros will play a role in the Bethe ansatz constraints.

\subsection{New family of QES potentials: polynomial deformations}
Let us consider the exactly solvable, SUSY transformed harmonic oscillator. Polynomial deformations of it break the exact solvability and introduce new model parameters. Under certain conditions on the parameters, the polynomial deformed potentials are QES.  

The general form of polynomial deformed QES potential with state-adding pseudo-Hermite polynomial can be written as
\begin{equation}\label{GR New QES potential}
    V=\frac{1}{2}\bigg[\sum_{k=1}^{\mu}A_k r^{2k}+g\frac{d^2}{dr^2}\ln{{\cal W}({\cal H}_{m_1},{\cal H}_{m_2},\cdots,{\cal H}_{m_N})}\bigg],\quad \mu\geq\text{deg}({\cal W}_N)-1=t-1,
\end{equation}
where $A_k$ are constant coefficients and $\mu$ is appropriate upper boundary integer so that the deformed potential is QES. In the follows we restrict to odd $\mu$, so that $\frac{\mu+1}{2}$ is an integer and the polynomial gauge factor in (\ref{GR wavefunction}) is well defined. Since $t=\deg({\cal W}_N)$ is even, the minimal choice $\mu=t-1$ is indeed odd, however, the value-selection rule by itself does not force every admissible deformation degree $\mu$ must be odd. The corresponding Schr\"odinger equation in the spherical coordinate is
\begin{equation}\label{GR Schrodinger equation}
\begin{split}
        \bigg[-\frac{d^2}{dr^2}+\frac{l(l+1)}{r^2}+\sum_{k=1}^{\mu}A_k r^{2k}+g\frac{d^2}{dr^2}\ln{{\cal W}({\cal H}_{m_1},{\cal H}_{m_2},\cdots,{\cal H}_{m_N})}\bigg]\psi(r)=2E\psi(r). 
\end{split}
\end{equation}
Write the wavefunction as the form
\begin{equation}\label{GR wavefunction}
    \psi(r)=r^{l+1}{\cal W}^{\nu}_{N}\exp[\omega(r)]\,f(r), \qquad \omega(r)=\sum_{p=1}^{\frac{\mu+1}{2}}B_p r^{2p}, \qquad \nu=\frac{1}{2}\bigg(1-\sqrt{1-4g}\bigg). 
\end{equation}
where ${\cal W}_N={\cal W}({\cal H}_{m_1},{\cal H}_{m_2},\cdots,{\cal H}_{m_N})$ and $f(r)$ is a polynomial of $r$. Substituting (\ref{GR wavefunction}) into (\ref{GR Schrodinger equation}), we obtain the ODEs
\begin{equation}\label{GR ODE}
    \begin{split}
      r{\cal W}_N  f''(r)+2\bigg[{(l+1)}{\cal W}_N+\nu r {{\cal W}_N'}+r\omega'(r){\cal W}_N\bigg]f'(r)+\bigg[2\nu\bigg(r\omega'(r)+{l+1}\bigg){{\cal W}_N'}+(\nu-g)r{{\cal W}_N''}\\
    +\bigg(2(l+1)\omega'(r)+r\omega''(r)+2Er\bigg) {\cal W}_N+ r{\cal W}_N\bigg(
      \omega'(r)^2-\sum_{k=1}^{\mu}A_k r^{2k}\bigg)\bigg]f(r)=0,
    \end{split}
\end{equation}
For ODE (\ref{GR ODE}) to have exact (polynomial) solutions, the degree of the polynomial in front of $f(r)$ must be less than the degree of the polynomial in front of $f'(r)$ by one.  This is achieved by choosing the coefficients $B_p$ in the gauge factor such that the monomial terms in the deformation of degree greater than $\mu/2$  are canceled by the higher degree terms in $(\omega'(r))^2$. This gives the relation that determines $B_p$ from $A_k$ with $k\geq (\mu+1)/2$ by matching the coefficients in front of $r^{2k}$,
\begin{equation}\label{GR Ak and Bp}
    \sum_{\substack{1\leq p,q\leq\frac{\mu+1}{2} \\ p+q=k+1}}4pqB_pB_q=A_k,\qquad k=\frac{\mu+1}{2},\frac{\mu+3}{2},\cdots,\mu.
\end{equation}
Here we present some solutions for $B_p$ and use $\sigma=\pm 1$ to represent the square roots branches. 
\begin{equation}
\begin{split}
        B_{\frac{\mu+1}{2}}=\frac{\sigma\sqrt{A_{\mu}}}{\mu+1},\quad
    B_{\frac{\mu-1}{2}}=\frac{A_{\mu-1}}{2\sigma(\mu-1)\sqrt{A_{\mu}}},\quad B_{\frac{\mu-3}{2}}=\frac{A_{\mu-2}-\frac{A_{\mu-1}^2}{4A_{\mu}}}{2\sigma(\mu-3)\sqrt{A_{\mu}}},\\B_{\frac{\mu-5}{2}}=\frac{1}{2\sigma(\mu-5)\sqrt{A_{\mu}}}\bigg[A_{\mu-3}-\frac{A_{\mu-1}}{2A_{\mu}}\bigg(A_{\mu-2}-\frac{A_{\mu-1}^2}{4A_{\mu}}\bigg)\bigg],\quad \cdots,
\end{split}
\end{equation}
In the following applications, for real sector with $A_\mu>0$, we choose the branch $\sigma=-1$ to make sure the wavefunctions admit square integrability. These results can be summarized in terms of the following recursive relations.
\begin{equation}
    \begin{split}
        B_{\frac{\mu+1}{2}-\epsilon}=&\frac{1}{4(\frac{\mu+1}{2}-\epsilon)q_0}\left(A_{\mu-\epsilon}-\sum_{\lambda=1}^{\epsilon-1}q_{\lambda}q_{\epsilon-\lambda}\right),\\ q_{\lambda}=&2\left(\frac{\mu+1}{2}-\lambda\right)B_{\frac{\mu+1}{2}-\lambda},\quad \epsilon=1,2,\cdots,\frac{\mu-1}{2}.
    \end{split}
\end{equation}

Recall, that $t=$deg$({\cal W}_N)$ and that $t$ is an even integer according to the value-selection rule of the $N$th-order SUSY. Now introduce the polynomials of $r$ of degrees $t+1, \mu+t+1,\mu+t$, respectively,
\begin{equation}\label{GR general form of ODE coefficients}
    \begin{split}
        X(r)=&r {\cal W}_N,\qquad Y(r)=2\bigg[(l+1){\cal W}_N+\nu r {\cal W}_N'+r\omega'(r){\cal W}_N\bigg],\\
        Z(r)=&2\nu \bigg(r\omega'(r)+l+1\bigg){\cal W}_N'(r)+r(\nu-g){\cal W}''_N+\bigg(2(l+1) \omega'(r)+r\omega''(r)\bigg){\cal W}_N\\
        &+{\cal W}_N\sum_{k=1}^{\frac{\mu-1}{2}}\left( \sum_{p+q=k+1}4 pq B_pB_q-A_k\right) r^{2k+1}+2Er{\cal W}_N. 
    \end{split}
\end{equation}
Then the ODE (\ref{GR ODE}) can be written in the form,
\begin{equation}\label{GR general form of ODE}
    \bigg[X(r)\frac{d^2}{dr^2}+Y(r)\frac{d}{dr}+Z(r)\bigg]f(r)=0.
\end{equation}
(\ref{GR general form of ODE}) 
is the ODE corresponding to the gauge-transformed Schr\"odinger equation with the new polynomial deformed QES potentials. It contains many free model parameters, the $A_k$'s, and point out that the degrees of polynomial terms depend on the degree of the Wronskian ($t$) as well as the degree of the deformations ($\mu$).

We are interested in polynomial solutions of the ODE of $n$ degree and express them in their factored form as 
\begin{equation}\label{BAE f(z)}
    f(r)=\prod_{i=1}^n(r-r_i), \qquad f(r) \equiv 1 \quad \text{for} \quad n=0,
\end{equation}
where $r_i$ are the roots of the polynomial. Substituting (\ref{BAE f(z)}) into (\ref{GR general form of ODE}) and evaluating the resulting equation at $r=r_i$, we obtain the so-called Bethe ansatz equations satisfied by the roots  \cite{zhang2012},
\begin{equation}\label{BAE}
    \sum_{j\neq i}^n \frac{2}{r_i-r_j}+\frac{Y(r_i)}{X(r_i)}=0,\qquad i=1,2,\ldots, n,
\end{equation}
or explicitly,
\begin{equation}\label{BAE polynomial deformation}
    \sum_{j \neq i}^n \frac{1}{r_i-r_j}+\bigg[\frac{l+1}{r_i}+\nu \frac{{\cal W}'_N (r_i)}{{\cal W}_N (r_i)}+\omega'(r_i)\bigg]=0,\qquad i=1,2,\ldots, n,
\end{equation}
together with certain constraints obeyed by the model parameters. 

Due to the general structure of the QES potential, the parameter constraints can not be written in a closed form in general. In the following, we present some relations which can be used to determine the energy and the model constraints. 

We expand (\ref{BAE f(z)}) in powers of $r$ in terms of symmetric polynomials $e_\ell$ of the Bethe roots $r_1, r_2, \ldots, r_n$,
\begin{equation}\label{GR symmetric polynomial}
\begin{split}
        f(r)=\sum_{s=0}^n(-1)^{n-s}e_{n-s}r^s.
\end{split}
\end{equation}
Then the coefficients $e_\ell$ are given by \cite{quesne2018quasi}
\begin{equation}\label{GR symmetric polynomial coefficient}
    \begin{split}
        &e_0 \equiv 1,\\
        &e_\ell \equiv e_\ell(r_1,r_2,\cdots,r_n)=\sum_{1\leq i_1<i_2<\cdots<i_l\leq n}r_{i_1}r_{i_2}\cdots r_{i_l}, \quad \ell=1,2,\cdots,n. 
    \end{split}
\end{equation}

We write (\ref{GR general form of ODE coefficients}) as
\begin{align}
    X(r)=\sum^{t+1}_{k=0}\alpha_k r^k, \qquad
        Y(r)=\sum^{\mu+t+1}_{k=0}\beta_k r^k, \qquad Z(r)=\sum_{k=0}^{\mu+t} \gamma_k r^k, 
\end{align}
where $\alpha_k, \beta_k, \gamma_k$ are coefficients related to the model parameters through (\ref{GR general form of ODE coefficients}).
Substituting (\ref{GR symmetric polynomial}) into (\ref{GR general form of ODE}), we obtain
\begin{equation}\label{recurrence relation polynomial deformation}
\begin{split}
    \sum^{t+1}_{k=0}\sum_{s=0}^{n-2}(-1)^{n-s-2}(s+2)(s+1)\alpha_ke_{n-s-2}r^{s+k}+\sum^{\mu+t+1}_{k=0}\sum_{s=0}^{n-1}(-1)^{n-s-1}\beta_k(s+1)e_{n-s-1}r^{s+k}\\+\sum_{k=0}^{\mu+t}\sum_{s=0}^n(-1)^{n-s} \gamma_ke_{n-s}r^{s+k}=0.
\end{split}
\end{equation}
Since the terms $r^d$ in (\ref{recurrence relation polynomial deformation}) are linearly independent, the parameter constraints are obtained by setting the coefficients of each monomial degree in (\ref{recurrence relation polynomial deformation}) equal to zero. Thus, all terms that contribute to the same final degree must be combined. 

We fix the target monomial $r^{n+q}$, where $q=0,1,\cdots,\mu+t$ and collect all contributions to that degree. For $0\leq q\leq t-1$, the contributions from all terms in $Xf''$, $Yf'$ and $Zf$ give
\begin{equation}\label{pd general parameter constrain 1}
    \begin{split}
        \sum_{p=0}^{\text{min}[n-2,t-q-1]} (-1)^p (n-p)(n-p-1)\alpha_{q+p+2} e_p+\sum_{p=0}^{\text{min}[n-1, \mu+t-q]}(-1)^p (n-p)\beta_{q+p+1}e_p\\+\sum_{p=0}^{\text{min}[n,\mu+t-q]}(-1)^p \gamma_{q+p}e_p=0,
    \end{split}
\end{equation}
For the higher degree sectors $t\leq q\leq \mu+t$, the contributions only come from $Yf'$ and $Zf$. 
\begin{equation}\label{pd general parameter constrain 2}
    \sum_{p=0}^{\text{min}[n-1, \mu+t-q]}(-1)^p (n-p)\beta_{q+p+1}e_p+\sum_{p=0}^{\text{min}[n,\mu+t-q]}(-1)^p \gamma_{q+p}e_p=0. 
\end{equation}

Energy follows directly from the (\ref{pd general parameter constrain 2}) with $q=t+1$. We first write $N$th-order Wronskian in a polynomial form.
\[
    {\cal W}_N=\sum_{i=0}^{t/2} w_i r^{2i},
\]
The closed form of energy is 
\begin{equation}\label{pd general energy}
    \begin{split}
        E=-\frac{1}{2w_{t/2}}\bigg[{\Gamma}_{t+1}+\sum_{s=0}^{n-1}\bigg((-1)^{n-s-1}\beta_{t+n-s+1}(s+1)e_{n-s-1}+(-1)^{n-s}\gamma_{t+n-s+1}e_{n-s}\bigg)\bigg],
    \end{split}
\end{equation}
where 
\[
    w_{t/2}=2^{\sum_{i=1}^N m_i} \prod_{1 \leq i<j \leq N} (m_j-m_i),
\]
where ${\Gamma}_{t+1}$ is the remainder of $\gamma_{t+1}$ after separating $E$ from it. 

Setting to zero the coefficients in front of $r^k $ with $k\geq n$ in (\ref{recurrence relation polynomial deformation}) leads to algebraic equations which give the general expressions of constraints for the model parameters of the QES potentials.

\subsection{New family of QES potentials: rational deformations}

Rational potentials play an important role in the study of ES quantum systems \cite{GRANDATI20112074, Grandati2013,Odake2013}. Rational deformations can be performed on ES potentials via SUSY transformations, leading to new ES models with EOP structures (see, e.g. \cite{Ho2011} \cite{Ho201110.1143}). In this section, we are interested in rational deformations which break the exact solvability and result in QES systems in which only finite wavefunctions and part of spectrum can be solved analytically.  We will construct new QES potentials based on the state-adding technique.

The general form of rational deformed QES potentials through state-adding of pseudo-Hermite polynomial can be expressed as 
\begin{equation}\label{RD New QES potential}
    V=\frac{1}{2}\bigg[\sum_{k=1}^{\mu}c_k r^{2k}+\sum_{i=1}^{\tau}\frac{a_ir^{2i}}{{\cal W}_N}+\sum_{j=1}^{\kappa}\frac{b_j r^{2j}}{{\cal W}_N^2}+g\frac{d^2}{dr^2}\ln{{\cal W}({\cal H}_{m_1},{\cal H}_{m_2},\cdots,{\cal H}_{m_N})}\bigg], 
\end{equation}
where $a_i, b_j$ and $c_k$ are constant model parameters associated with the deformation, $\mu, \tau$ are appropriate upper bound integers of the deformation terms, $\kappa=\text{deg}({\cal W}_N)-1=t-1$, and ${\cal W}_N={\cal W}({\cal H}_{m_1},{\cal H}_{m_2},\cdots,{\cal H}_{m_N})$. The corresponding Schr\"odinger equation is
\begin{equation}\label{RD Schrodinger equation}
\begin{split}
        \bigg[-\frac{d^2}{dr^2}+\frac{l(l+1)}{r^2}+\sum_{k=1}^{\mu}c_k r^{2k}+\sum_{i=1}^{\tau}\frac{a_ir^{2i}}{{\cal W}_N}+\sum_{j=1}^{\kappa}\frac{b_j r^{2j}}{{\cal W}_N^2}+g\frac{d^2}{dr^2}\ln{{\cal W}({\cal H}_{m_1},{\cal H}_{m_2},\cdots,{\cal H}_{m_N})}\bigg]\psi(r)=2E\psi(r). 
\end{split}
\end{equation}
Set
\begin{equation}\label{RD wavefunction}
    \psi(r)=r^{l+p+1}\,{\cal W}_N^\nu \,f(r)\exp[{\omega(r)}] , \qquad \omega(r)=\sum_{q=1}^{M}B_q r^{2q},\qquad \nu=\frac{1}{2}\left(1\pm\sqrt{1-4g+\frac{4b_{\kappa } }{ t^2 w_{t/2}^2}}\right), 
\end{equation}
where $w_{t/2}$ is the coefficient of the highest degree term in ${\cal W}_N$, $p=0,1$ correspond to the even and odd sectors of the system, respectively, and $f(r)$ is a polynomial of $r$. $M$ is certain upper bound integer to be determined by the highest effective degrees in the first two summations in (\ref{RD New QES potential}). Substituting (\ref{RD wavefunction}) into (\ref{RD Schrodinger equation}), we obtain the gauge-transformed ODE for $f(r)$,
\begin{equation}\label{rational general ODE0}
    \begin{split}
        f''(r)+2\left(\nu\frac{{\cal W}_N'}{{\cal W}_N}+\omega'(r)+\frac{l+p+1}{r} \right)f'(r)+\bigg[2\nu \left(\omega'(r)+\frac{l+p+1}{r}\right)\frac{{\cal W}_N'}{{\cal W}_N}+2\omega'(r)\frac{l+p+1}{r}\\
        \qquad\qquad+\frac{[\nu(\nu-1)+g]({\cal W}_N')^2-\sum_{j=1}^{\kappa}b_j r^{2j}}{{\cal W}_N^2}+(\nu-g)\frac{{\cal W}_N''}{{\cal W}_N}+\omega''(r)+(\omega'(r))^2\\
        \qquad\qquad -\sum_{i=1}^{\tau}\frac{a_ir^{2i}}{{\cal W}_N}-\sum_{k=1}^{\mu}c_k r^{2k}-\frac{2p(l+1)}{r^2}+2E\bigg]f(r)=0. 
    \end{split}
\end{equation}
With the choice of $\kappa=t-1$, the deformation terms provide sufficient number of $b_j$ to match the terms in  $[\nu(\nu-1)+g]({\cal W}_N')^2$ so that the contributions from the two terms proportional to $1/{\cal W}_N^2$ cancel and the ODE is simplified to
\begin{equation}\label{rational general ODE1}
    \begin{split}
       r^2{\cal W}_N f''(r)+2r\bigg[\nu r{{\cal W}_N'}+\left(r\omega'(r)+l+p+1\right){\cal W}_N \bigg]f'(r)
       +\bigg[2\nu r\left(r\omega'(r)+{l+p+1}\right){{\cal W}_N'}\\
       +(\nu-g)r^2{{\cal W}_N''}+r\left[2(l+p+1)\omega'(r)+r\omega''(r)\right]{\cal W}_N-2p(l+1){\cal W}_N\\+r^2\bigg((\omega'(r))^2{\cal W}_N
        -{\cal W}_N\sum_{k=1}^{\mu}c_k r^{2k}-\sum_{i=1}^{\tau}{a_ir^{2i}}\bigg)+2E{\cal W}_Nr^2\bigg]f(r)=0. 
    \end{split}
\end{equation}

We choose the coefficients $B_q$ and integer $M$ in the gauge-factor $\omega(r)$ such that the terms proportional to $r^s$ with $s>2M+t$ in $\displaystyle (\omega'(r))^2{\cal W}_N
        -{\cal W}_N\sum_{k=1}^{\mu}c_k r^{2k}-\sum_{i=1}^{\tau}{a_ir^{2i}}$ cancel.
Write 
\[
\omega'(r)^2=\sum_{j=1}^{2M-1}d_jr^{2j}=\sum_{\substack{1\leq p,q\leq M\\p+q=j+1}}4pqB_pB_q r^{2(p+q)-2}.
\]
Then the cancellation gives the constraint for $M$,
\begin{equation}
4M-2=\text{max}[2\mu,2\tau-t]
\end{equation}
and the recursive relation for $B_q$,
\begin{equation}
    \begin{split}
        B_{M-\epsilon}=\frac{1}{4(M-\epsilon)q_0}\left(d_{2M-1-\epsilon}-\sum_{\lambda=1}^{\epsilon-1}q_{\lambda}q_{\epsilon-\lambda}\right),\\
        q_{\lambda}=2(M-\lambda)B_{M-\lambda},\quad \epsilon=1,2,\cdots,M-1. 
    \end{split}
\end{equation}

It then follows that similarly to the polynomial deformation case, after the cancellation the ODE (\ref{rational general ODE1}) can be written in the standard form,
\begin{equation}\label{rational general ODE}
    \bigg[\tilde{X}(r)\frac{d^2}{dr^2}+\tilde{Y}(r)\frac{d}{dr}+\tilde{Z}(r)\bigg]f(r)=0.
\end{equation}
where
\begin{equation}
    \begin{split}\label{2.36}
        \tilde{X}(r)=&r^2{\cal W}_N,\qquad \tilde{Y}(r)=2\left[\nu r^2 {\cal W}'_N+ r^2\omega'(r){\cal W}_N+(l+p+1)r {\cal W}_N\right], \\
        \tilde{Z}(r)=&2\nu r \left(r\,\omega'(r)+{l+p+1}\right){\cal W}'_N+r\bigg(2(l+p+1)\omega'(r)+r\omega''(r)\bigg){\cal W}_N\\
        &+(\nu-g)r^2 {\cal W}''_N+\sum_{k=1}^{\frac{t}{2}+M-1}\left(q_k-p_k\right)r^{2k+2}-2p(l+1){\cal W}_N+2Er^2 {\cal W}_N
    \end{split}
\end{equation}
are polynomials in $r$ of degrees $t+2, 2M+t+1, 2M+t$, respectively, and 
$p_k$ are defined from the relation 
\[
{\cal W}_N\sum_{k=1}^{\mu}c_k r^{2k}+\sum_{i=1}^{\tau}a_i r^{2i}=\sum_{i=1}^{(t+D)/2}p_ir^{2i},\quad D=\text{max}[2\mu,2\tau-t]
\]

The ODE (\ref{rational general ODE}) is the general form of the gauge-transformed Schr\"odinger equation with the new rational QES potential. An important feature of this choice of deformation is that it would not introduce new singularities into the system on the real line. This is because the rational terms are initially obtainable from Wronskians of the pseudo Hermite polynomial, they ensure that the corresponding system is within the framework of Darboux-Crum (state adding) and that no further singularities are added.
Note that, similar to the polynomial deformation case, the rational QES deformation introduces many free parameters into the system ODE and thus quasi-exact solvability of the system is fulfilled only for parameters which satisfy specific constraints. 

In what follows, we present a general analysis of solutions and constraints of the system via a combination of Bethe ansatz and recurrence relations.  In particular, we seek closed-form solutions in the form (\ref{BAE f(z)}). Substituting the ansatz (\ref{BAE f(z)}) into the ODE (\ref{rational general ODE}), we find that the roots $r_i$ satisfy the following Bethe ansatz equations, 
\begin{equation}\label{BAE rational deformation}
    \sum_{j\neq i}^n \frac{1}{r_i-r_j}+2\left[\nu\frac{{\cal W}_N'(r_i)}{{\cal W}_N(r_i)}+\omega'(r_i)+\frac{l+p+1}{r_i} \right]=0,\quad i=1,2,\ldots, n. 
\end{equation}
To determine the energy and constraints for the model parameters, similarly to the polynomial deformation case, we expend (\ref{BAE f(z)}) in terms of symmetrical polynomials $e_\ell$ of the Bethe roots in the form as given in
(\ref{GR symmetric polynomial}), and write
\begin{align}\label{2.38}
    \tilde{X}(r)=\sum_{k=0}^{t+2}\tilde{\alpha}_kr^k,\qquad
    \tilde{Y}(r)=\sum_{k=0}^{2M+t+1}\tilde{\beta}_k r^k, \qquad
    \tilde{Z}(r)=\sum_{k=0}^{2M+t}\tilde{\gamma}_k r^k,
\end{align}
where $\tilde{\alpha}_k, \tilde{\beta}_k, \tilde{\gamma}_k$ are coefficients related to the model parameters of the rational deformation through (\ref{2.36}). Substituting (\ref{GR symmetric polynomial}) into (\ref{rational general ODE}), we obtain
\begin{equation}\label{rational 3 terms recurrence relation}
    \begin{split}
        \sum^{t+2}_{k=0}\sum_{s=0}^{n-2}\tilde{\alpha}_k(-1)^{n-s-2}(s+2)(s+1)e_{n-s-2}r^{s+k}+\sum^{t+2M+1}_{k=0}\sum_{s=0}^{n-1}\tilde{\beta}_k(-1)^{n-s-1}(s+1)e_{n-s-1}r^{s+k}\\+\sum_{k=0}^{t+2M}\sum_{s=0}^n(-1)^{n-s}e_{n-s} \tilde{\gamma}_k r^{s+k}=0. 
    \end{split}
\end{equation}
The parameter constraints are obtained following a procedure similar to the previous subsection. For $0\leq q\leq t$, the contributions to $r^{n+q}$ give
\begin{equation}\label{rd general parameter constrain 1}
    \begin{split}
        \sum_{p=0}^{\text{min}[n-2,t-q]}(-1)^p (n-p)(n-p-1)\tilde{\alpha}_{q+p+2} e_p+\sum_{p=0}^{\text{min}[n-1,2M+t-q]}(-1)^p (n-p)\tilde{\beta}_{q+p+1}e_p\\+\sum_{p=0}^{\text{min}[n,2M+t-q]}(-1)^p \tilde{\gamma}_{q+p}e_p=0
    \end{split}
\end{equation}
For the higher degree with $t+1\leq q\leq 2M+t$, we have
\begin{equation}\label{rd general parameter constrain 2}
    \begin{split}
        \sum_{p=0}^{\text{min}[n-1,2M+t-q]}(-1)^p (n-p)\tilde{\beta}_{q+p+1}e_p+\sum_{p=0}^{\text{min}[n,2M+t-q]}(-1)^p \tilde{\gamma}_{q+p}e_p=0. 
    \end{split}
\end{equation}
The explict form of energy can be obtained from (\ref{rd general parameter constrain 2}) with $q=t+2$.  
\begin{equation}\label{rd general energy}
    \begin{split}
        E=-\frac{1}{2w_{t/2}}\bigg[\tilde{\Gamma}_{t+2}+\sum_{s=0}^{n-1}\bigg((-1)^{n-s-1}\tilde{\beta}_{t+n-s+2}(s+1)e_{n-s-1}+(-1)^{n-s}\tilde{\gamma}_{t+n-s+2}e_{n-s}\bigg)\bigg],
    \end{split}
\end{equation}
where 
\[
    w_{t/2}=2^{\sum_{i=1}^N m_i} \prod_{1 \leq i<j \leq N} (m_j-m_i),
\]
where $\tilde{\Gamma}_{t+2}$ is the remainder of the $\tilde{\gamma}_{t+2}$ after separating $E$ from it.

\section{Applications}\label{Polynomial deformation}

Families of polynomial and rational QES deformations were introduced in Sections 2.3 and 2.4, in which formulas for the associated parameter constraints and functional Bethe ansatz methods were also given. In this section,  as applications we give in-dept analysis of different important examples. 

\subsection{QES Polynomial deformations of 1st order SUSY systems related to $X_m$ EOPs}
\subsubsection{Codimension 2 (i.e. $m=2$) case}
In this subsection, we consider the first member (i.e. $m=2$ case) in the family of systems (\ref{GR New QES potential}). The corresponding QES potential reads \cite{agboola2014}
\begin{equation}\label{CE QES potential m=2}
    V_2=\frac{1}{2}A_1 r^2-2g\frac{2r^2-1}{(2r^2+1)^2},
\end{equation}
Making the change of the wavefunction 
\begin{equation}
    \psi(r)=r^{l+1}{\cal H}_2^{\nu}\,f(r)\,\exp[{\omega(r)}], \qquad \omega(r)=B_1 r^2, \quad \nu=\frac{1}{2}\bigg(1-\sqrt{1-4g}\bigg), 
\end{equation}
we obtain the transformed ODE for $f(r)$,
\begin{equation}
    \begin{split}
        f''(r)+2\bigg(\frac{l+1}{r}+\frac{4\nu r}{2r^2+1}+2B_1 r\bigg)f'(r)+4\bigg[\frac{2\nu (2B_1 r^2+l+1)+\nu-g}{2r^2+1}\\+\frac{B_1}{2} (2l+3)
        +B_1^2 r^2-\frac{A_1}{4}r^2+\frac{E}{2}\bigg]f(r)=0.
    \end{split}
\end{equation}
Changing the variable $z=r^2$ and setting $B_1=-\frac{\sqrt{A_1}}{2}$, allows us to set the ODE in standard form (\ref{GR general form of ODE}), which is 
\begin{equation}\label{CE QES ODE z m=2}
    \begin{split}
        2z(2z+1)f''(z)+ \left[-4 \sqrt{A_1} z^2+ \left(-2 \sqrt{A_1}+4 l+8 \nu +6\right)z+2 l+3\right]f'(z)\\
        +\bigg[ \left(2E-\sqrt{A_1}(4\nu+2l+3)\right)z-\frac{\sqrt{A_1}}{2}(2l+3)+ E-2 g+2\nu(2l+3)\bigg]f(z)=0. 
    \end{split}
\end{equation}

We look for exact solutions of the form
 \begin{equation}\label{CE QES polynomial solutions m=2}
 \begin{split}
        f(z)=\prod_{i=1}^n(z-z_i),\qquad 
        f(z) \equiv 1 ~ {\rm when} ~ n=0.
 \end{split}
 \end{equation}
Expressing $f(z)$ in terms of the symmetric polynomials $e_\ell$ as in (\ref{GR symmetric polynomial}), we have from (\ref{CE QES ODE z m=2}),
\begin{equation}
    \begin{split}
        \left[2E-\sqrt{A_1}\bigg(4n+4\nu+2l+3\bigg)\right]e_0z^{n+1}+\bigg\{[-2E+\sqrt{A_1}(4n+4\nu+2l-1)]e_1\\
        +\bigg[E-\frac{\sqrt{A_1}}{2} (2 l+4 n+3)-2 g+2 \nu  (2 l+4 n+3)+2 n (2 l+2 n+1)\bigg]e_0\bigg\}z^n\\
        -\sum_{k=0}^{n-1}(-1)^{n-k}\bigg\{
         \bigg[2E-\sqrt{A_1}\bigg(4k+4\nu+2l-1\bigg)\bigg]e_{n-k+1}
        +\bigg[ E-\frac{\sqrt{A_1}}{2} (2 l+4 k+3)\\-2 g+2 \nu  (2 l+4 k+3)+2 k (2 l+2k+3)\bigg]e_{n-k}
        +(k+1)(2k+2l+3)e_{n-k-1}\bigg\}z^{k}=0.
    \end{split}
\end{equation}
Setting the coefficients of $z^{n+1}$ and $z^n$  respectively to zero  gives the energy and the constraint of the model parameters of the system, 
\begin{align}
    &\begin{aligned}\label{CE QES SUSY1 m=2 parameter constrain E}
        E_n=\sqrt{A_1}\bigg(2n+2\nu +l+\frac{3}{2}\bigg), 
    \end{aligned}\\
    &\begin{aligned}\label{CE QES SUSY1 m=2 parameter constrain other}
        \sqrt{A_1} \left(\nu -2 \sum_{i=1}^n z_i\right)+\nu  (2 l +3)-g+n (2n+2 l+4 \nu +1)=0,
    \end{aligned}
\end{align}
where the roots $z_i$ are determined from the Bethe ansatz equations
\begin{equation}\label{CE QES SUSY1 m=2 BAE}
    \sum_{j \neq i}^n \frac{2}{z_i-z_j}+\frac{-2 \sqrt{A_1} z_i^2+\left(- \sqrt{A_1}+2 l+4 \nu +3\right)z_i+(2 l+3)/2}{z_i(2z_i+1)}=0,\qquad i=1,2,\ldots,n. 
\end{equation}

\subsubsection{Codimension 4 (i.e. $m=4$) case}
For $m=4$, the QES potential is
\begin{equation}\label{CE SUSY1 m=4 V}
    V=\frac{1}{2} \left(A_3 r^6+A_2 r^4+A_1 r^2\right)-\frac{4 g \left(8 r^6+12 r^4+18 r^2-9\right)}{\left[4 (r^2+3) r^2+3\right]^2}
\end{equation}
Making the transformation for the wavefunction $\psi(r)$ in the Schr\"odinger equation with the above potential,
\begin{equation}
        \psi(r)=r^{l+1}{\cal H}_4^{\nu}(r)e^{\omega(r)}f(r),\quad \omega(r)=\sum_{p=1}^{2}{ B}_p r^{2p},\quad \nu=\frac{1}{2}\left(1-\sqrt{1-4g}\right),
\end{equation}
we obtain the gauge-transformed ODE
\begin{equation}\label{CE SUSY1 m=4 ODE r}
    \begin{split}
f''(r)+2 \bigg[4 B_2 r^3+2 B_1 r+\frac{l+1}{r}+\frac{8 \nu  \big(2 r^2+3\big) r}{4 \big(r^2+3\big) r^2+3}\bigg]f'(r)+\bigg[-A_3 r^6-A_2 r^4-A_1 r^2\\+4 (l+1) \big(2 B_2 r^2+B_1\big)+\frac{16 \nu  \big(2 r^2+3\big) \big(4 B_2 r^4+2 B_1 r^2+l+1\big)}{4 \big(r^2+3\big) r^2+3}+4 \big(2 B_2 r^3+B_1 r\big)^2\\+12 B_2 r^2+2 B_1+2 E-\frac{24 \big(2 r^2+1\big) (g-\nu )}{4 \big(r^2+3\big) r^2+3}\bigg]f(r)=0
    \end{split}
\end{equation}
We choose the coefficients of $B_p$ to be
\begin{equation}\label{CE SUSY1 m=4 gauge factor}
    B_1=-\frac{A_2}{4 \sqrt{A_3}},\quad B_2=-\frac{1}{4} \sqrt{A_3}
\end{equation}
so that the terms proportional to $r^6$ and $r^4$ in (\ref{CE SUSY1 m=4 ODE r}) cancel, respectively, and the ODE is now in the standard form (\ref{GR general form of ODE}). Now, making the change of variable $z=r^2$ and in terms of symmetrical polynomials $e_\ell$, we can show that the ODE (\ref{CE SUSY1 m=4 ODE r}) becomes
\begin{equation}\label{CE QES SUSY1 m=4 symmetric ODE}
    \begin{split}
    Q_2(z)\sum_{s=0}^{n-2}(-1)^{n-s-2}(s+2)(s+1)e_{n-s-2}z^s+Q_1(z)\sum_{s=0}^{n-1}(-1)^{n-s-1}(s+1)e_{n-s-1}z^s\\+Q_0(z)\sum_{s=0}^{n}(-1)^{n-s}e_{n-s}z^s=0.
    \end{split}
\end{equation}
where $Q_2(z), Q_1(z), Q_0(z)$ are given by
\begin{equation}\label{m=4 Q2}
     Q_2(z)=16 z^3+48 z^2+12 z,
\end{equation}
\begin{equation}\label{m=4 Q1}
\begin{split}
         Q_1(z)=-16 \sqrt{A_3} z^4-\left(\frac{8 A_2}{\sqrt{A_3}}+48 \sqrt{A_3}\right) z^3+\bigg(-12 \sqrt{A_3}-\frac{24 A_2}{\sqrt{A_3}}+8 (2 l+3)+64 \nu \bigg)z^2\\+\bigg[24 (2 l+3)+96 \nu -\frac{6 A_2}{\sqrt{A_3}}\bigg]z+6 (2 l+3),
\end{split}
\end{equation}
\begin{equation}\label{m=4 Q0}
\begin{split}
         Q_0(z)=\bigg[-4 \sqrt{A_3} (2 l+5)-32 \sqrt{A_3} \nu +\frac{A_2^2}{A_3}-4 A_1\bigg]z^3+\bigg[-\frac{2 A_2 (2 l+3)}{\sqrt{A_3}}-12 \sqrt{A_3} (2 l+5)-\frac{16 A_2 \nu }{\sqrt{A_3}}\\-48 \sqrt{A_3} \nu +\frac{3 A_2^2}{A_3}-12 A_1+8 E\bigg]z^2+\bigg[-\frac{6 A_2 (2 l+3)}{\sqrt{A_3}}-3 \sqrt{A_3} (2 l+5)-\frac{24 A_2 \nu }{\sqrt{A_3}}\\+\frac{3 A_2^2}{4 A_3}-3 A_1+24 E-48 g+32 l \nu +80 \nu \bigg]z-\frac{3 A_2 (2 l+3)}{2 \sqrt{A_3}}+6 E-24 g+48 l \nu +72 \nu.
\end{split}
\end{equation}
As in the $m=2$ case, quasi-exact solvability  requires the vanishing of the polynomial coefficients in front of $z^k$ with $k\geq n$. This gives the energy 
\begin{equation}
    \begin{split}
        E=2 \sqrt{A_3} \sum_{i=1}^n z^{i}+\frac{A_2 (2 l+8 \nu +4 n+3)}{4 \sqrt{A_3}}+\frac{3}{2} \sqrt{A_3} (2 l+4 \nu +4 n+5)-\frac{3 A_2^2}{8 A_3}+\frac{3 A_1}{2}
    \end{split}
\end{equation}
and the constraints for the model parameters (whose explicit expressions are omitted here due to their involved form). The roots $z_i$ are determined by the Bethe ansatz equations
\begin{equation}\label{CE SUSY1 m=4 BAE}
\begin{split}
        \sum_{j\neq i}^n \frac{2}{z_i-z_j}+\frac{8\nu(2z_i+3)}{4 z_i (z_i+3)+3} +\frac{2l+3}{2z_i}-\frac{A_2}{2\sqrt{A_3}}-\sqrt{A_3}z_i=0, \quad i=1,\ldots,n.
\end{split}
\end{equation}

\subsubsection{Codimension 6 (i.e. $m=6$) case}
In this subsection, we look at system with larger codimension 6. As seen from the previous subsection, the complexity increases dramatically  as the rational term gets more involved. 

The QES potential is \label{CE SUSY1 m=6 V}
\begin{equation}
    V=\frac{1}{2} \bigg(A_5 r^{10}+A_4 r^8+A_3 r^6+A_2 r^4+A_1 r^2\bigg)-6g\frac{450 r^2+8 \big(4 r^6+30 r^4+90 r^2+75\big) r^4-225}{\big(8 r^6+60 r^4+90 r^2+15\big)^2}
\end{equation}
After gauge transformation with gauge-factor $\displaystyle \omega(r)=\sum_{p=1}^{3}{ B}_p r^{2p}$ to the wavefunction $\psi(r)$, the corresponding equation Scr\"odinger becomes 
\begin{equation}\label{CE SUSY1 m=6 ODE r}
    \begin{split}
        f''(r)+2 \bigg[6 B_3 r^5+4 B_2 r^3+2 B_1 r+\frac{l+1}{r}+\frac{12 \nu  \big(4 \big(r^2+5\big) r^2+15\big) r}{8 r^6+60 r^4+90 r^2+15}\bigg]f'(r)+\bigg[-A_5 r^{10}-A_4 r^8-A_3 r^6\\-A_2 r^4-A_1 r^2+4 (l+1) \big(3 B_3 r^4+2 B_2 r^2+B_1\big)+\frac{24 \nu  \big(4 \big(r^2+5\big) r^2+15\big) \big(6 B_3 r^6+4 B_2 r^4+2 B_1 r^2+l+1\big)}{8 r^6+60 r^4+90 r^2+15}\\+30 B_3 r^4+12 B_2 r^2+4 \big(3 B_3 r^5+2 B_2 r^3+B_1 r\big)^2+2 B_1+2 E-\frac{60 \big(4 \big(r^2+3\big) r^2+3\big) (g-\nu )}{8 r^6+60 r^4+90 r^2+15}\bigg]f(r)=0. 
    \end{split}
\end{equation}
We choose the following coefficients $B_p$ in the gauge-factor $\omega(r)$
\begin{equation}\label{CE SUSY1 m=6 gauge facotr}
    B_1=\frac{A_4^2-4 A_3 A_5}{16 A_5^{3/2}},\quad B_2=-\frac{A_4}{8 \sqrt{A_5}},\quad B_3=-\frac{1}{6} \sqrt{A_5}
\end{equation}
so that the terms involving $r^{10}, r^8, r^6$ in (\ref{CE SUSY1 m=6 ODE r}) cancel, respectively, and the ODE is in the standard form (\ref{GR general form of ODE}). Furthermore, we make the change of variable $z=r^2$ and use the symmetric polynomials $e_\ell$ to bring the (\ref{CE SUSY1 m=6 ODE r}) to the form
\begin{equation}\label{CE QES SUSY1 m=6 symmetric ODE}
    \begin{split}
        Q_2(z)\sum_{s=0}^{n-2}(-1)^{n-s-2}(s+2)(s+1)e_{n-s-2}z^s+Q_1(z)\sum_{s=0}^{n-1}(-1)^{n-s-1}(s+1)e_{n-s-1}z^s\\+Q_0(Z)\sum_{s=0}^{n}(-1)^{n-s}e_{n-s}z^s=0,
    \end{split}
\end{equation}
where 
\begin{equation}\label{m=6 Q2}
     Q_2(z)=32 z^4+240 z^3+360 z^2+60 z,
\end{equation}
\begin{equation}\label{m=6 Q1}
    \begin{split}
         Q_1(z)=-32 \sqrt{A_5} z^6-\bigg(\frac{16 A_4}{\sqrt{A_5}}+240 \sqrt{A_5}\bigg) z^5+\bigg(\frac{4 A_4^2}{A_5^{3/2}}-\frac{120 A_4}{\sqrt{A_5}}-360 \sqrt{A_5}-\frac{16 A_3}{\sqrt{A_5}}\bigg) z^4+\bigg(\frac{30 A_4^2}{A_5^{3/2}}\\-\frac{180 A_4}{\sqrt{A_5}}-60 \sqrt{A_5}-\frac{120 A_3}{\sqrt{A_5}}+16 (2 l+3)+192 \nu \bigg)z^3 + \bigg(\frac{45 A_4^2}{A_5^{3/2}}-\frac{30 A_4}{\sqrt{A_5}}-\frac{180 A_3}{\sqrt{A_5}}\\+120 (2 l+3)+960 \nu \bigg)z^2+\bigg(\frac{15 A_4^2}{2 A_5^{3/2}}-\frac{30 A_3}{\sqrt{A_5}}+180 (2 l+3)+720 \nu \bigg)z+30 (2 l+3),
    \end{split}
\end{equation}
\begin{equation}\label{m=6 Q0}
    \begin{split}
         Q_0(z)=\bigg(-\frac{A_4^3}{A_5^2}+\frac{4 A_3 A_4}{A_5}-8 A_2-8 (2 l+7) \sqrt{A_5}-96 \nu  \sqrt{A_5}\bigg) z^5+\bigg(\frac{A_4^4}{8 A_5^3}-\frac{15 A_4^3}{2 A_5^2}-\frac{A_3 A_4^2}{A_5^2}+\frac{30 A_3 A_4}{A_5}\\-\frac{4 (2 l+5) A_4}{\sqrt{A_5}}-\frac{48 \nu  A_4}{\sqrt{A_5}}-8 A_1-60 A_2-60 (2 l+7) \sqrt{A_5}-480 \nu  \sqrt{A_5}+\frac{2 A_3^2}{A_5}\bigg) z^4\\+\bigg(\frac{15 A_4^4}{16 A_5^3}-\frac{45 A_4^3}{4 A_5^2}-\frac{15 A_3 A_4^2}{2 A_5^2}+\frac{(2 l+3) A_4^2}{A_5^{3/2}}+\frac{12 \nu  A_4^2}{A_5^{3/2}}+\frac{45 A_3 A_4}{A_5}-\frac{30 (2 l+5) A_4}{\sqrt{A_5}}-\frac{240 \nu  A_4}{\sqrt{A_5}}\\+16 E-60 A_1-90 A_2-90 (2 l+7) \sqrt{A_5}-360 \nu  \sqrt{A_5}+\frac{15 A_3^2}{A_5}-\frac{4 (2 l+3) A_3}{\sqrt{A_5}}-\frac{48 \nu  A_3}{\sqrt{A_5}}\bigg) z^3\\+\bigg(\frac{45 A_4^4}{32 A_5^3}-\frac{15 A_4^3}{8 A_5^2}-\frac{45 A_3 A_4^2}{4 A_5^2}+\frac{60 \nu  A_4^2}{A_5^{3/2}}+\frac{15 (2 l+3) A_4^2}{2 A_5^{3/2}}+\frac{15 A_3 A_4}{2 A_5}-\frac{45 (2 l+5) A_4}{\sqrt{A_5}}\\-\frac{180 \nu  A_4}{\sqrt{A_5}}+120 E-240 g+96 l \nu +336 \nu -90 A_1-15 A_2-15 (2 l+7) \sqrt{A_5}+\frac{45 A_3^2}{2 A_5}\\-\frac{30 (2 l+3) A_3}{\sqrt{A_5}}-\frac{240 \nu  A_3}{\sqrt{A_5}}\bigg) z^2+\bigg(\frac{15 A_4^4}{64 A_5^3}-\frac{15 A_3 A_4^2}{8 A_5^2}+\frac{45 \nu  A_4^2}{A_5^{3/2}}+\frac{45 (2 l+3) A_4^2}{4 A_5^{3/2}}-\frac{15 (2 l+5) A_4}{2 \sqrt{A_5}}\\+180 E-720 g+480 l \nu +1200 \nu -15 A_1+\frac{15 A_3^2}{4 A_5}-\frac{45 (2 l+3) A_3}{\sqrt{A_5}}-\frac{180 \nu  A_3}{\sqrt{A_5}}\bigg) z\\+30 E-180 g+360 l \nu +540 \nu +\frac{15 (2 l+3) A_4^2}{8 A_5^{3/2}}-\frac{15 (2 l+3) A_3}{2 \sqrt{A_5}}.
    \end{split}
\end{equation}

The energy and constraints for the model parameters can be obtained by requiring the vanishing of the polynomial coefficients in front of $r^k$ with $k\geq n$ in (\ref{CE QES SUSY1 m=6 symmetric ODE}). The calculations are similar to those for the $m=2$ case and we will not repeat the process here. In the following, we only present the energy expression
 \begin{equation}
    \begin{split}
        E=2 \sqrt{A_5}\left[\left(\sum_{i=1}^n z_i\right)^2-2\sum_{1\leq i<j\leq n} z_i z_j\right]+\left(\frac{A_4}{\sqrt{A_5}}+15\sqrt{A_5}\right)\sum_{i=1}^n z_i-\frac{A_4^2 (2 l+12 \nu +4 n+3)}{16 A_5^{3/2}}-\frac{15 A_4^4}{256 A_5^3}\\+\frac{15 (2 A_3+3 A_4) A_4^2}{64 A_5^2}+\frac{15}{8} (2 A_1+3 A_2)-\frac{15 A_3 (A_3+3 A_4)}{16 A_5}+\frac{15 A_4 (2 l+8 \nu +4 n+5)}{8 \sqrt{A_5}}\\+\frac{A_3 (4 l+24 \nu +8 n+6)}{8 \sqrt{A_5}}+\frac{45}{8} \sqrt{A_5} (2 l+4 \nu +4 n+7). 
    \end{split}
\end{equation}
The roots $z_i$ are determined by the Bethe ansatz equations 
\begin{equation}\label{CE SUSY1 m=6 BAE}
    \begin{split}
        \sum_{j\neq i}^n\frac{2}{z_i-z_j}+\frac{12\nu \big( 4z_i^2+20z_i+15\big)}{8 z_i^3+60 z_i^2+90 z_i+15}+\frac{2l+3}{2z_i}-\sqrt{A_5} z_i^2-\frac{A_4 z_i+A_3}{2\sqrt{A_5}}+\frac{A_4^2}{ 8A_5^{3/2}}=0, \quad i=1,\ldots, n.
    \end{split}
\end{equation}

\subsubsection{Numeric results of allowed model parameters for the above three cases}

\begin{figure}[H]
  \centering
  \begin{subfigure}{0.34\linewidth}
    \centering
    \includegraphics[width=\linewidth]{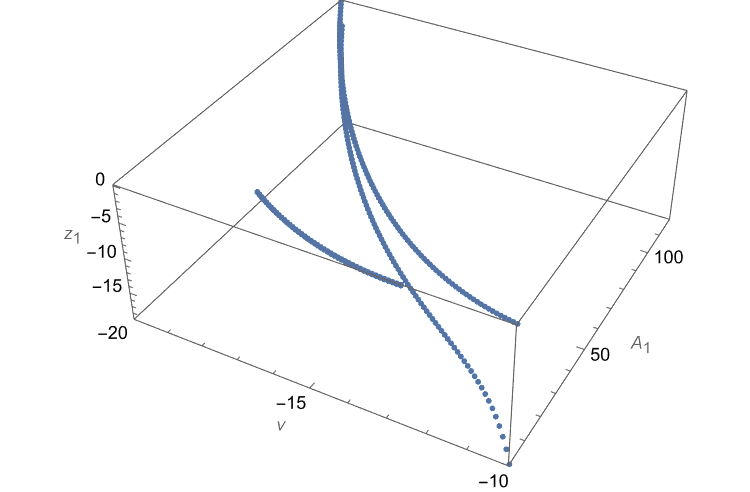}
    \caption{Allowed parameters for $m=2$}
    \label{f 1st_m=2}
  \end{subfigure}
  \hfill
  \begin{subfigure}{0.32\linewidth}
    \centering
    \includegraphics[width=\linewidth]{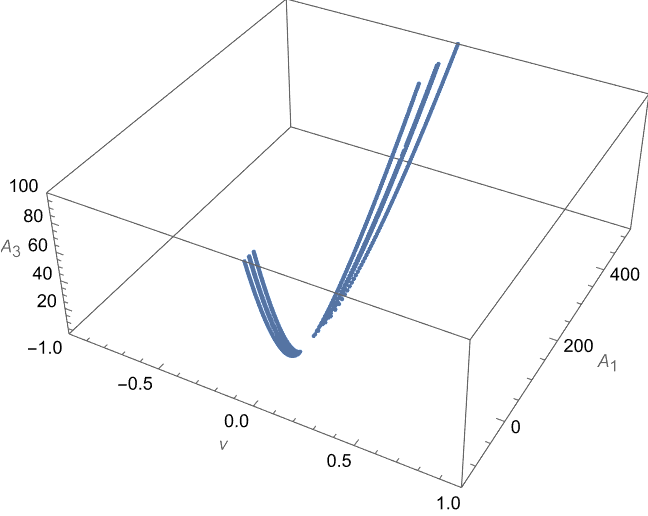}
    \caption{Allowed parameters for $m=4$}
    \label{f 1st_m=4}
  \end{subfigure}
    \hfill
  \begin{subfigure}{0.32\linewidth}
    \centering
    \includegraphics[width=\linewidth]{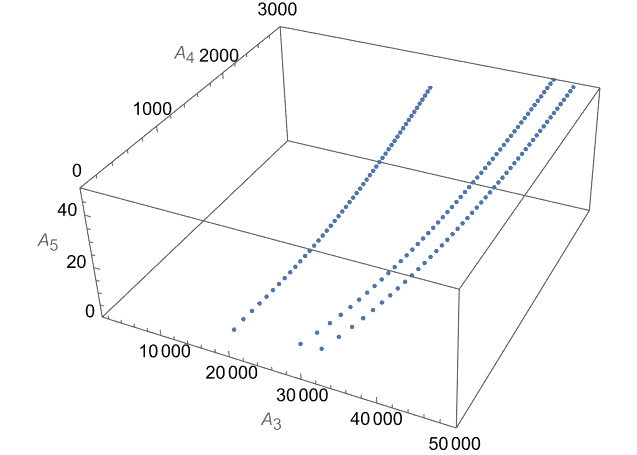}
    \caption{Allowed parameters for $m=6$}
    \label{f 1st_m=6}
  \end{subfigure}
  \caption{Allowed parameters in the QES potentials of the 1st order SUSY transformation. }
  \label{fig 1st}
\end{figure}

Fig.(\ref{fig 1st}) below  provides allowed values for the model parameters in the parameter space for QES potentials based on the 1st order SUSY transformation with $m=2$, $m=4$ and $m=6$. They give the corresponding parameter constraints and demonstrate quasi-exactly solvability of the polynomial deformations and thus the existence of QES potentials. We list the numerical results of the first order SUSY cases here. For the higher order SUSY cases, we provide the associated numerical results at appendix \ref{appendix A}. 

We set the parameters $n=2, l=3$ and change the value $\nu$ from -20 to -10 and plot the admitted values of $A_1$ and one of the Bethe roots $z_1$. There are 3 lines in the parameter space in Fig.(\ref{f 1st_m=2}), which indicates there are 3 independent solutions for $n=2$. The structure of the ODE satisfy the QES classification by Turbiner\cite{turbiner1988quasi}. Thus, the ODE has hidden $sl(2)$ symmetry. For $m=4$ case, we fix $n=2, l=3$ and change the $A_3$ from 1 to 50 to avoid the singularities. There are a couple of lines in Fig.(\ref{f 1st_m=4}) and each line represents an eigenfunction of the ODE. The structure of the ODE beyond would then not be among the $sl(2)$ algebraic classification of Quasi exactly solvable systems from $m=4$. Therefore, the hidden $sl(2)$ symmetry is broken and the number of the solutions does not satisfy the $n+1$ rule. For $m=6$ case, we reduce to $n=1, l=3$ and change the value of $A_5$ from 1 to 50, because of the complicated of ODE. The lines provide the allowed parameters in the ODE and indicate the QES potential exists solutions in Fig.(\ref{f 1st_m=6}). The approach can be applied on systems with larger codimension, the Bethe Ansatz and the parameter constraints are more involved, but from computational point of view the approach can be implemented. However, the Bethe Ansatz are taking more complicated form and additional parameter constraints appear. Our calculation have also been implemented with $n=2$ as it is non-trivial, however higher excited states can be solved to higher order. It is known that at certain limit Bethe Ansatz are difficult to solve, here we have also the parameter constraints that need to be solve all together and supplementary constraints are we want real energy and parameter solutions to only look at certain physical case.

\subsection{ Quasi-exactly solvable deformations of exceptional Hermite systems}
\label{sec:qes-exceptional-hermite}

We consider rational deformations of the harmonic oscillator associated with
exceptional Hermite systems.  Our focus is on families depending continuously
on a deformation parameter $\epsilon$ and reducing, at a distinguished value
$\epsilon=\epsilon_0$, to an exactly solvable exceptional Hermite system.  The
examples below illustrate the behavior of the algebraic solutions in this
limit.

\paragraph{Degree deformations associated with a degree two seed}
\label{subsec:degree-two-seed}

We begin with rational deformations associated with a degree two seed
solution.  In this case, the deformed system admits polynomial solutions for
every degree $n$.  We display the first four solutions, corresponding to $
  n=0,1,2,3.$ 
  
Although all four solutions are well defined away from the exactly solvable
limit, their limiting behavior is different.  The solutions with $n=0$ and
$n=3$ reduce to physical eigenfunctions belonging to the exceptional Hermite
system.  In contrast, the solutions with $n=1$ and $n=2$ reduce to nonphysical
solutions of the limiting system.

For each $n$, let $P_n(x;\epsilon)$ denote the polynomial factor and write the
corresponding quasi-polynomial eigenfunction as
\[
  \psi_n(x;\epsilon)
  = \mu_2(x;\epsilon)P_n(x;\epsilon),
\]
where $\mu_2$ is the appropriate gauge factor for the degree two seed.

\subsubsection*{The solution with $n=0$}

\begin{align}
 V_0(x;\epsilon) &= \epsilon^2 x^6
+(\epsilon^2+2\epsilon)x^4
+\left(\frac{\epsilon^2}{4}+2\epsilon+1\right)x^2
-\frac{\epsilon}{2}+3
-\frac{16}{(2x^2+1)^2}
+\frac{8}{2x^2+1},\\
 E_0(\epsilon)   &= \epsilon /2, \\
  \psi_0(x;\epsilon) &= 
-\frac{\exp\left(
-\frac{x^2}{4}\left(\epsilon x^2+\epsilon+2\right)
\right)}
{2(2x^2+1)}.
 \end{align}

\subsubsection*{The solution with $n=1$}

\begin{align}
  V_1(x;\epsilon) &= \epsilon^2 x^6
+(\epsilon^2-2\epsilon)x^4
+\left(\frac{\epsilon^2}{4}-2\epsilon+1\right)x^2
-\frac{3\epsilon}{2}-1
+\frac{8(2x^2-1)}{(2x^2+1)^2}, \\
  E_1(\epsilon)   &= \frac{3 \epsilon}{2}+4, \\
  \psi_1(x;\epsilon) &= -\frac{x}{2(2x^2+1)}
\exp\left(
-\frac{x^2\left(\epsilon x^2+\epsilon-2\right)}{4}
\right).
\end{align}


\subsubsection*{The solution with $n=2$}

\begin{align}
  V_2(x;\epsilon) &= \epsilon^2x^6
+2\epsilon^2x^4
+\left(\frac{5\epsilon^2}{4}+1\right)x^2
+2-2\epsilon
-\left(\epsilon x^4+\epsilon x^2+\frac12\right)
\sqrt{\epsilon^2+12\epsilon+4}+\frac{8(2x^2-1)}{(2x^2+1)^2}, \\
  E_2(\epsilon)   &= 2\epsilon + 1 + \frac{1}{2}\sqrt{\epsilon^2 + 12\epsilon + 4}, \\
  \psi_2(x;\epsilon) &= \frac{
-4\epsilon x^2
+\sqrt{\epsilon^2+12\epsilon+4}
-\epsilon-2
}{
8\epsilon(2x^2+1)
}
\exp\left[
\frac{x^2}{4}
\left(
-\epsilon x^2
+\sqrt{\epsilon^2+12\epsilon+4}
-2\epsilon
\right)
\right].
\end{align}

\subsubsection*{The solution with $n=3$}

\begin{align}
  V_3(x;\epsilon) &= \epsilon^2 x^6
+2\epsilon^2 x^4
+\left(\frac{5\epsilon^2}{4}+1\right)x^2
-2-4\epsilon
+\left(\epsilon x^4+\epsilon x^2-\frac12\right)
\sqrt{\epsilon^2+20\epsilon+4}+\frac{8(2x^2-1)}{(2x^2+1)^2}, \\
  E_3(\epsilon)   &= 4\epsilon + 5 + \frac{1}{2}\sqrt{\epsilon^2 + 20\epsilon + 4}, \\
  \psi_3(x;\epsilon) &= -\frac{
x\left(
4\epsilon x^2
+\sqrt{\epsilon^2+20\epsilon+4}
+\epsilon-2
\right)
}{
8\epsilon(2x^2+1)
}
\exp\left[
-\frac{x^2}{4}
\left(
\epsilon x^2
+\sqrt{\epsilon^2+20\epsilon+4}
+2\epsilon
\right)
\right].
\end{align}

\begin{figure}[htbp]
    \centering

    \includegraphics[width=0.48\textwidth]{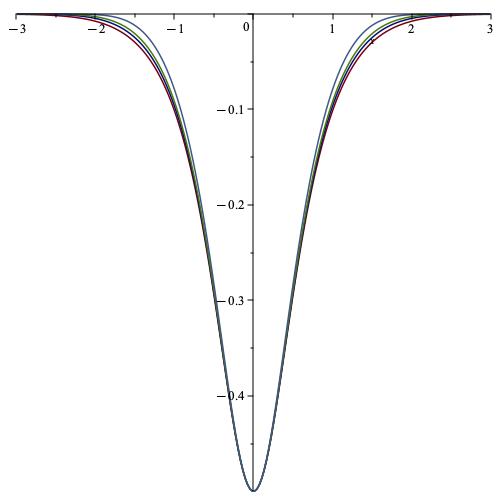}
    \hfill
    \includegraphics[width=0.48\textwidth]{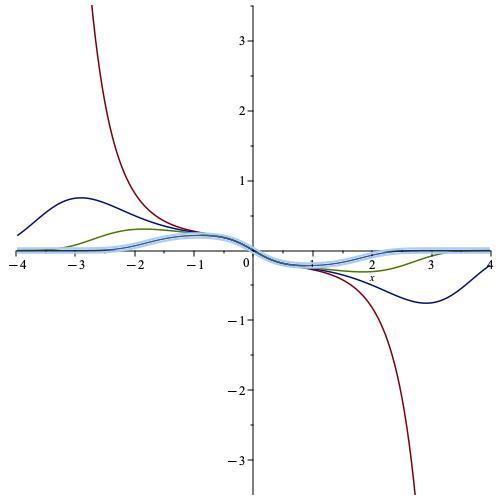}

    \vspace{0.3cm}

    \includegraphics[width=0.48\textwidth]{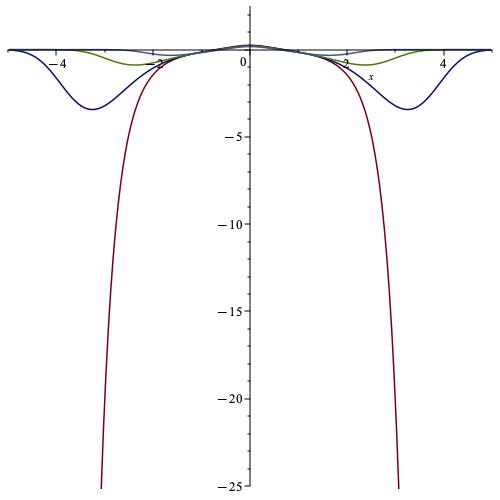}
    \hfill
    \includegraphics[width=0.48\textwidth]{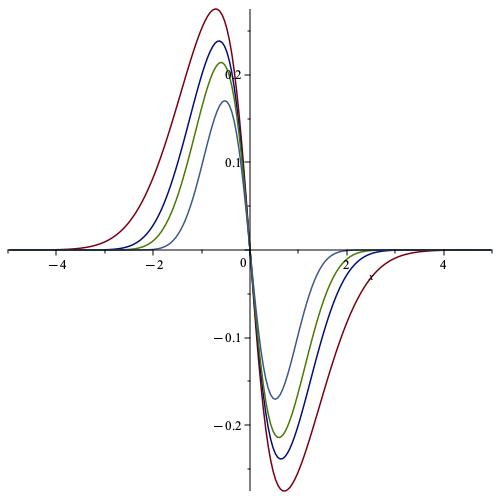}

    \caption{The first four wave functions, $\psi_0,\psi_1,\psi_2,\psi_3$,
    for $\epsilon=0,\,0.1,\,0.2,$ and $0.5$.}
    \label{fig:m2-wavefunctions}
\end{figure}

\paragraph{Rational deformations associated with a degree four seed}
\label{subsec:degree-four-seed}

For a degree four seed, polynomial deformations yield polynomial solutions
only at particular values of $\epsilon$.  To obtain a continuously
parameterized family that approaches the exceptional Hermite limit, we instead
consider rational deformations.  The resulting systems admit quasi-polynomial
solutions for general $\epsilon$. Notice however that the linear solution does not reduce to the corresponding exactly
solvable rational oscillator as $\epsilon\to\epsilon_0$.

\subsubsection*{Constant solution}

\begin{align}
  V_{0}(x;\epsilon) &= \epsilon^2 x^6
+\left(5\epsilon^2+2\epsilon\right)x^4
+\left(\frac{25}{4}\epsilon^2+10\epsilon+1\right)x^2
+\frac{11}{2}\epsilon+7\\
&\qquad + \frac{24\epsilon}{4x^4 + 12x^2 + 3}
+\frac{16\left(8x^6+12x^4+15x^2-9\right)}
{\left(4x^4+12x^2+3\right)^2}, \\
  E_{0}(\epsilon)   &= -\frac{11 \epsilon}{2}, \\
  \psi_{0}(x;\epsilon)
    &= \frac{1}{4\left(4x^4+12x^2+3\right)}
\exp\left[
-\frac{x^2}{4}\left(\epsilon x^2+5\epsilon+2\right)
\right].
\end{align}

\subsubsection*{Linear solution}

\begin{align}
  V_{1}(x;\epsilon) &= \epsilon^2 x^6
+\left(5\epsilon^2+\frac{2}{3}\epsilon\right)x^4
+\left(\frac{25}{4}\epsilon^2+\frac{14}{3}\epsilon+\frac{1}{9}\right)x^2
+\frac{\epsilon}{2}+\frac{5}{3}
\\
&\qquad + \qquad + \frac{8(3\epsilon + 4)}{4x^4 + 12x^2 + 3}
+\frac{16\left(8x^6+12x^4+15x^2-9\right)}
{\left(4x^4+12x^2+3\right)^2}, \\
  E_{1}(\epsilon)   &= -\frac{\epsilon}{2}+\frac{16}{3}, \\
  \psi_{1}(x;\epsilon)
    &= \frac{x}{4\left(4x^4+12x^2+3\right)}
\exp\left[
-\frac{x^2}{12}\left(3\epsilon x^2+15\epsilon+2\right)
\right].
\end{align}

\begin{figure}[htbp]
    \centering

    \includegraphics[width=0.48\textwidth]{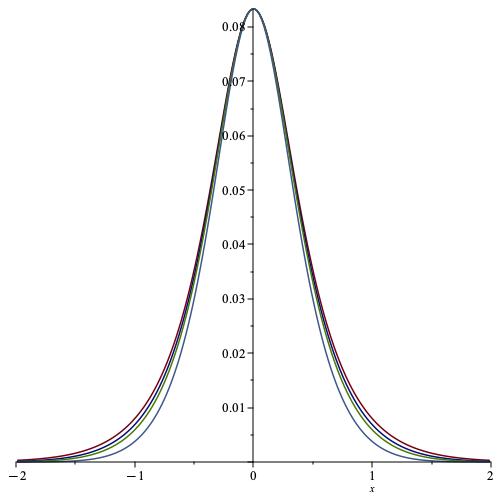}
    \hfill
    \includegraphics[width=0.48\textwidth]{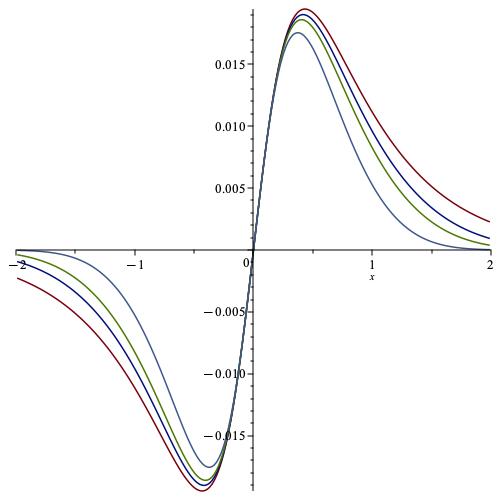}

    \caption{The degree $0$ and degree $1$ solutions, respectively,
    for rational deformations associated with a degree $4$ seed function,
    for $\epsilon=0,\,0.1,\,0.2,$ and $0.5$.}
    \label{fig:m4-rational-deformations}
\end{figure}

\subsection{QES polynomial deformation of higher-order SUSY systems related to multi-indexed EOPs}\label{higher order SUSY m1=2 m2=3}

As the initial exactly-solvable system is created via higher order SUSYQM, there are more than one gaps and the codimensions are needed to take into account the different gaps. The associated multi-indexed EOPs of Hermite type are usually denoted as $X_{m_1,...,m_{N}}$, where $m_i$ refer to the gaps. In this section, we demonstrate that higher order QES extensions exist for appropriate anharmonic deformations.

In the case of second order SUSY transformation with $m_1=2, m_2=3$, the new QES potential takes the form 
\begin{equation}\label{CE SUSY2 m1=2 m2=3 QES potential}
    V=\frac{1}{2}\alpha r^2+\frac{1}{2}\beta r^4+\frac{1}{2}\gamma r^6-8g\frac{4r^6-9r^2}{(3+4r^4)^2}
\end{equation}
We make a gauge-transformation to the wavefunction $\psi(r)$,  
\begin{equation}
    \psi(r)=2 r^{l+1} (m_2{\cal H}_{m_1}{\cal H}_{m_2-1}-m_1{\cal H}_{m_1-1}{\cal H}_{m_2}) (r)e^{\omega(r)}f(r), 
\end{equation}
where $f(r)$ is a polynomial, $\omega(r)=a r^2+br^4$ and $\nu=\frac{1}{2}\left(1-\sqrt{1-4g}\right)$. 

Then the Schr\"odinger equation is given by the ODE 
\begin{equation}\label{CE SUSY2 m1=2 m2=3 ODE r}
    \begin{split}
        f''(r)+2\bigg[\frac{8 a r^6+6 a r^2+4 l r^4+3 l+16 \nu  r^4+4 r^4+3}{r (4 r^4+3)}+4 b r^3\bigg]f'(r)+ \frac{1}{3+4r^4}\bigg[r^6 (-4 \alpha-3 \gamma\\+16 (b (3 b+2 l+8 \nu +5)+a^2))+r^4 (8 a (6 b+2 l+8 \nu +3)-3 \beta)+r^2 (-3 \alpha+12 (b (2 l+5)+a^2)\\+16 (2 l+5) \nu )-4 r^{10} (\gamma-16 b^2)-4 r^8 (\beta-16 a b)+12 a l+18 a+E (8 r^4+6)-48 g r^2\bigg]f(r)=0
    \end{split}
\end{equation}
We choose $a=-\frac{\beta}{4\sqrt{\gamma}}, b=-\frac{\sqrt{\gamma}}{4}$ so that the terms involving $r^6$ and $r^4$ in the above ODE cancel, respectively. Making a variable change $z=r^2$, the gauge-transformed ODE becomes 
\begin{equation}\label{CE SUSY2 m1=2 m2=3 QES transformed ODE z}
    \begin{split}
        &(16z^3+12z)f''(z)+\bigg[-16\sqrt{\gamma}z^4-\frac{8\beta}{\sqrt{\gamma}}z^3+\big(16l+64\nu-12\sqrt{\gamma}+24\big)z^2-\frac{6\beta}{\sqrt{\gamma}}z+12l+18\bigg]f'(z)\\
        &+\bigg[z^2 \bigg(-\frac{(4l+16\nu+6) \beta }{\sqrt{\gamma}}+8 E\bigg)+z \bigg(-6 \sqrt{\gamma} l+\frac{3 \beta^2}{4 \gamma}-3 \alpha-15 \sqrt{\gamma}-48 g+32 l \nu +80 \nu \bigg)\\
        &+z^3 \bigg(-8 \sqrt{\gamma} l-32 \sqrt{\gamma} \nu +\frac{\beta^2}{\gamma}-4 \alpha-20 \sqrt{\gamma}\bigg)-\frac{3 \beta l}{\sqrt{\gamma}}-\frac{9 \beta}{2 \sqrt{\gamma}}+6 E\bigg]f(z)=0
    \end{split}
\end{equation}
Using the symmetric polynomials $e_\ell$ for $f(z)$, we have
\begin{equation}\label{CE QES SUSY2 m1=2 m2=3 symmetric ODE}
    \begin{split}
        (16z^3+12z)\sum_{s=0}^{n-2}(-1)^{n-s-2}(s+2)(s+1)e_{n-s-2}z^s+\bigg[-16\sqrt{\gamma}z^4-\frac{8\beta}{\sqrt{\gamma}}z^3+\big(16l+64\nu-12\sqrt{\gamma}\\+24\big)z^2-\frac{6\beta}{\sqrt{\gamma}}z+12l+18\bigg]\sum_{s=0}^{n-1}(-1)^{n-s-1}(m+1)e_{n-s-1}z^s
        +\bigg[z^2 \bigg(-\frac{(4l+16\nu+6) \beta }{\sqrt{\gamma}}+8 E\bigg)\\+z \bigg(-6 \sqrt{\gamma} l+\frac{3 \beta^2}{4 \gamma}-3 \alpha-15 \sqrt{\gamma}-48 g+32 l \nu +80 \nu \bigg)
    +z^3 \bigg(-8 \sqrt{\gamma} l-32 \sqrt{\gamma} \nu +\frac{\beta^2}{\gamma}-4 \alpha-20 \sqrt{\gamma}\bigg)\\-\frac{3 \beta l}{\sqrt{\gamma}}-\frac{9 \beta}{2 \sqrt{\gamma}}+6 E\bigg]\sum_{s=0}^{n}(-1)^{n-s}e_{n-s}z^s=0.
    \end{split}
\end{equation}
The closed form expression for the energy is
\begin{equation}
    \begin{split}
        E=2 \sqrt{\gamma} \sum_{i=1}^n z_i+\frac{\beta}{4 \sqrt{\gamma}} (2 l+8 \nu +4 n+3), 
    \end{split}
\end{equation}
where the roots satisfy the Bethe ansatz equations 
\begin{equation}
    \begin{split}
        \sum_{j\neq i}^2 \frac{2}{z_i-z_j}+\frac{-8\sqrt{\gamma}z_i^4-\frac{4\beta}{\sqrt{\gamma}}z_i^3+\big(8l+32\nu-6\sqrt{\gamma}+12\big)z_i^2-\frac{3\beta}{\sqrt{\gamma}}z_i+6l+9}{8z_i^3+6z_i}=0,\quad i=1,2,\ldots,n. 
    \end{split}
\end{equation}


\subsection{QES rational deformation of first order SUSYQM with $m=2$: $\tau=1$ case}

Based on the state-adding of SUSY for the harmonic oscillator (\ref{SUSY ES harmonic original pseudo-Hermite polynomial}), we construct a new QES potential for $m=2$. 
\begin{equation}\label{BD potential}
    V=\frac{1}{2}\left[cr^2+\frac{ar^2}{{\cal H}_2}+\frac{br^2}{{\cal H}_2^2}+g\left(\frac{{\cal H}_2''}{{\cal H}_2}-\left(\frac{{\cal H}_2'}{{\cal H}_2}\right)^2\right)\right]
\end{equation}
where $a,b,c,g$ are parameters. 
We make a gauge transformation to the wavefucntion
\begin{equation}\label{BD wavefunction}
    \psi(r)=r^{1+l+p}({\cal H}_2)^\delta \exp{\left(-\frac{\sqrt{c}}{2}r^2\right)}f(r),\quad \delta=\frac{1}{8}\left(4+\sqrt{16-64g+b}\right), \quad p=0,1, 
\end{equation}
where $p=0,1$ label the states in the even and odd sectors, respectively. 

Making change of variable $z=r^2$ in the corresponding Schr\"odinger equation, we obtain the gauge transformed ODE 
\begin{equation}\label{BD ODE}
    \begin{split}
        (16z^3+8z^2)f''(z)+\bigg[-16\sqrt{c}z^3+4\left(10+4l+4p+\sqrt{16-64g+b}-2\sqrt{c}\right)z^2+4(3+2l+2p)z\bigg]f'(z)\\+ \bigg\{\bigg[8E-a-2\sqrt{c}\left(4p+4l+10+\sqrt{16-64g+b}\right)\bigg]z^2\\+\bigg[4E-8g+8p(l+1)+(2l+3+2p)\bigg(4-2\sqrt{c}+\sqrt{16-64g+b}\bigg)\bigg]z+4p(l+1)\bigg\}f(z)=0.
    \end{split}
\end{equation}
The closed form expression of energy $E$ is given by
\begin{equation}
    \begin{split}
        E=\frac{1}{8} \left[a+2 \sqrt{c} \left(\sqrt{b-64 g+16}+4 l+8 n+4 p+10\right)\right],
    \end{split}
\end{equation}
and the zeros $z_i$ in the exact polynomial solutions of $f(z)$ are determined by the Bethe ansatz equations
\begin{equation}
    \begin{split}
        \sum_{j\neq i}^n\frac{2}{z_i-z_j}+\frac{-4\sqrt{c}\,z_i^2+\left(10+4l+4p+\sqrt{16-64g+b}-2\sqrt{c}\right)z_i+3+2l+2p}{2z_i(2z_i+1)}=0,\quad i=1,\ldots,n.
    \end{split}
\end{equation}

Substituting (2.20) into the ODE (3.39) and collecting the coefficients
with the same degree of $z$, we can find the constraints for the allowed model parameters. Their expressions are too long to be listed here. For demonstration, we provide some numeric results for the $n=2$ case. 
Fig. \ref{fig rd}(a) below provides the diagram for the allowed model parameters for $n=2$ case.
We fix $p=0, g=1, l=3$ and set $c$ from -20 to -10. The allowed values for the model parameters $a,b$ are represented by the blue lines, which correspond to the analytic solutions of the wavefunctions.

\subsection{QES rational deformation of first order SUSYQM with $m=2$: $\tau=2$ case}

The second rational deformation QES potential has the form
\begin{equation}\label{BD2 QES potential}
    V=\frac{1}{2}\left[cr^2+\frac{a_1 r^2+a_2 r^4}{{\cal H}_2}+\frac{br^2}{{\cal H}_2^2}+g\left(\frac{{\cal H}_2''}{{\cal H}_2}-\left(\frac{{\cal H}_2'}{{\cal H}_2}\right)^2\right)\right]
\end{equation}
where $a_1, a_2, b, c, g$ are parameters. 
We make the gauge transformation for the wavefunction $\psi(r)$, 
\begin{equation}\label{BD2 wavefunction}
     \psi(r)=r^{1+l+p}({\cal H}_2)^\delta \exp{\left(-\frac{\sqrt{a_2+4c}}{4}r^2\right)}f(r),\quad \delta=\frac{1}{8}\left(4+\sqrt{16-64g+b}\right), 
\end{equation}
where $p=0,1$ corresponds to even and odd sectors, respectively. 

Changing variable $z=r^2$ in the corresponding Schr\"odinger equation gives rise to the gauge transformed ODE
\begin{equation}\label{BD2 ODE 1}
    \begin{split}
        (16 z^3+8 z^2)f''(z)+\big[-8\sqrt{a_2+4c} z^3+4\left(10-\sqrt{a_2+4c} +\sqrt{16+b-64g}+4l+4p\right)z^2+(12\\+8l+8p)z\big]f'(z)+\bigg\{\left[\frac{a_2}{2}-a_1+8E-\sqrt{a_2+4c}\left(4l+4p+\sqrt{16+b-64g}+10\right)\right]z^2\\+\bigg[4E-8g+8p(l+1)+(2l+3+2p)\bigg(4-\sqrt{a_2+4c}+\sqrt{16-64g+b}\bigg)\bigg]z+4p(l+1)\bigg\}f(z)=0. 
    \end{split}
\end{equation}
The zeros of the exact polynomial solution $f(z)$ are determined by the Bethe ansatz equations
\begin{equation}
    \begin{split}
        \sum_{j\neq i}^n\frac{2}{z_i-z_j}+\frac{-4\sqrt{a_2+4c} z_i^2+\left(10-\sqrt{a_2+4c} +\sqrt{16+b-64g}+4l+4p\right)z_i+2l+3+2p}{2z_i(2z_i+1)}=0,\quad i=1,\ldots,n
    \end{split}
\end{equation}
and the closed form expression for the energy is given by
\begin{equation}
    \begin{split}
        E=\frac{1}{16} \left[2 a_1+2 \sqrt{a_2+4 c} \left(\sqrt{b-64 g+16}+4 l+8 n+4 p+10\right)-a_2\right].
    \end{split}
\end{equation}

\begin{figure}[H]
  \centering
  \begin{subfigure}{0.45\linewidth}
    \centering
    \includegraphics[width=\linewidth]{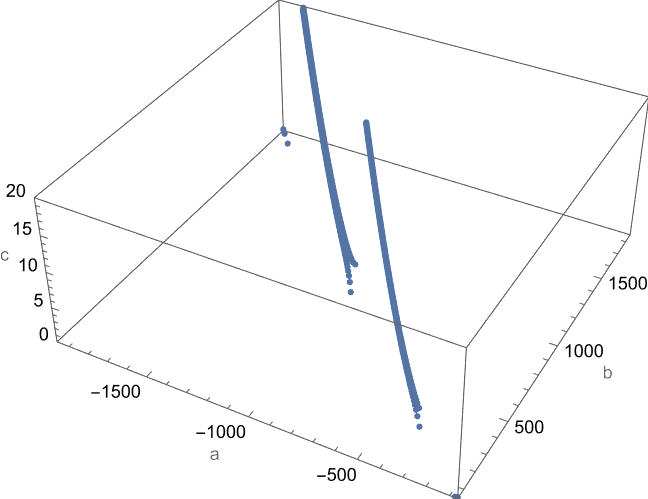}
    \caption{Allowed parameter values for the rational deformation 1}
    \label{f rd1}
  \end{subfigure}
  \hfill
  \begin{subfigure}{0.45\linewidth}
    \centering
    \includegraphics[width=\linewidth]{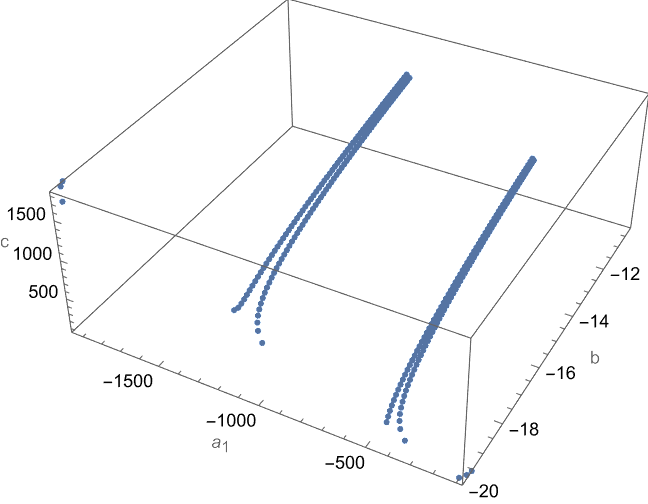}
    \caption{Allowed parameter values for the rational deformation 2}
    \label{f rd2}
  \end{subfigure}
  \caption{Allowed model parameters for the rational deformations. }
  \label{fig rd}
\end{figure}

Similar to the previous case, substituting (2.20) into the ODE (3.44) and collecting the coefficients
with the same degree of $z$, we can find the constraints for the allowed model parameters. We would not list their expressions here. For demonstration, we provide some numeric results for the $n=2$ case. 
Fig. \ref{fig rd}(b) below provides the diagram for the allowed model parameters for $n=2$ case.
We fix $p=0, g=1, l=3, c=5$ and set $a_2$ from -20 to -10. In this case, we plot the admitted values of the model parameters $a_1, b, c$ with blue lines, corresponding the analytic solutions.

\section{Conclusion}

One of the main results of this work is the proposal of a new framework to construct QES deformations of ES systems based on Darboux-Crum transformations and the functional Bethe ansatz. This approach is different from our recent work, in which supersymmetric transformations were applied directly to known QES systems to obtain new ones. The two ways of deforming ES systems are not equivalent as they lead to different QES models.

The new method enabled us to obtain families of anharmonic and rational deformations of ES systems related to exceptional orthogonal polynomials of Hermite type. We derived the constraints on model parameters required for quasi-exact solvability and the corresponding Bethe ansatz solutions of the deformed systems. We also presented the solution curves of the QES constraints in the parameter space determined using numerical techniques. 

It is interesting to investigate whether or not the large families of QES deformations obtained in this paper admit any hidden symmetry algebras. If any, they are possibly beyond Lie algebras.

\section*{Acknowledgment}

\noindent IM was supported by the Australian Research Council Future Fellowship FT180100099. SL was supported by a La Trobe Graduate Research Scholarship (LTGRS) and a La Trobe University Full-Fee Research Scholarship (LTFFRS).

\section*{Data availability}

\noindent The manuscript has no associated data.

\vskip.6in

\appendix

\section{Coefficients of gauge factors and some numeric results in higher order SUSY}\label{appendix A}
\vskip.1in
\noindent \underline{\bf \large 2nd-order SUSY with $m_1=4, m_2=5$}:
\vskip.1in
The QES potential has the form 
\begin{equation}\label{CE SUSY2 m1=4 m2=5 V}
    V=\sum_{i=1}^7 \frac{1}{2}A_kr^{2k}+g\frac{16  r^2 \left(16 r^8+72 r^4-135\right) \left[4 \left(r^2+5\right) r^2+15\right]}{\left[8 \left(2 r^4+8 r^2+15\right) r^4+45\right]^2}
\end{equation}
The corresponding gauge-transformed ODE is 
\begin{equation}\label{CE SUSY2 m1=4 m2=5 ODE r}
    \begin{split}
        f''(r)+2 \bigg[8 B_4 r^7+6 B_3 r^5+4 B_2 r^3+2 B_1 r+\frac{l+1}{r}+\frac{32 \nu  \left(4 \left(r^2+3\right) r^2+15\right) r^3}{8 \left(2 r^4+8 r^2+15\right) r^4+45}\bigg]f'(r) \\+\bigg[-A_7 r^{14}-A_6 r^{12}-A_5 r^{10}-A_4 r^8-A_3 r^6-A_2 r^4-A_1 r^2+4 (l+1) \big(4 B_4 r^6+3 B_3 r^4\\+2 B_2 r^2+B_1\big)+\frac{64 \nu  \left(4 \left(r^2+3\right) r^2+15\right) r^2 \left(8 B_4 r^8+6 B_3 r^6+4 B_2 r^4+2 B_1 r^2+l+1\right)}{8 \left(2 r^4+8 r^2+15\right) r^4+45}\\+56 B_4 r^6+30 B_3 r^4+12 B_2 r^2+4 \left(4 B_4 r^7+3 B_3 r^5+2 B_2 r^3+B_1 r\right)^2+2 B_1+2 E\\-\frac{32 \left(28 r^4+60 r^2+45\right) r^2 (g-\nu )}{8 \left(2 r^4+8 r^2+15\right) r^4+45}\bigg]f(r) =0
    \end{split}
\end{equation}
To cancel the higher order terms, the coefficients in the gauge factor  are chosen to be
\begin{equation}\label{CE SUSY2 m1=4 m2=5 gauge factors}
    \begin{split}
        B_4=-\frac{1}{8} \sqrt{A_7},\quad B_3=-\frac{A_6}{12 \sqrt{A_7}},\quad B_2=\frac{A_6^2-4 A_5 A_7}{32 A_7^{3/2}},\quad B_1=\frac{-A_6^3+4 A_5 A_7 A_6-8 A_4 A_7^2}{32 A_7^{5/2}}. 
    \end{split}
\end{equation}

\vskip.2in
\noindent \underline{\bf \large 3rd order SUSY with $m_1=2,m_2=3,m_3=4$}:
\vskip.1in
For this case, the QES potential is 
\begin{equation}\label{CE SUSY 3 QES potential m1=2 m2=3 m3=4}
    V=\sum_{i=1}^5\frac{1}{2}a_i r^{2i}-g\frac{192r^{10}-288r^8-288r^6-1296r^4+972r^2-162}{(8r^6-12r^4+18r^2+9)^2}. 
\end{equation}
The wavefunction is 
\begin{equation}\label{CE SUSY 3 wavefunction m1=2 m2=3 m3=4}
    \psi(r)=r^{l+1}{\cal W}_{(m_1,m_2,m_3)}^{\nu}(r)e^{\omega(r)}f(r),\quad \omega(r)=\sum_{p=1}^3{B}_p r^{2p},\quad \nu=\frac{1}{2}\bigg(1-\sqrt{1-4g}\bigg),
\end{equation}
the coefficients of the gauge factor are
\begin{equation}\label{CE SUSY 3 gauge factors coefficients}
    B_3=-\frac{\sqrt{A_5}}{6},\quad B_2=-\frac{A_4}{8 \sqrt{A_5}},\quad B_1=\frac{A_4^2-4 A_3 A_5}{16 A_5^{3/2}}. 
\end{equation}

\vskip.2in
\noindent \underline{\bf \large 4th order SUSY with $m_1=2, m_2=3, m_3=4, m_4=5$}:
\vskip.1in
For this case, the minimum top value of the polynomial $\sum_{k=1}^{\mu} A_k r^{2k}$ is $\mu_{min}=\sum_{i=1}^4m_i-7$. However, limited degrees of freedom, the numerical results cannot be solved easily. We choose a greater top value $u=\sum_{i=1}^4m_i-5$ to solve the QES potential based on 4th order SUSY. The QES potential is 
\begin{equation}\label{CE SUSY 4 QES potential m1=2 m2=3 m3=4 m4=5}
    V=\sum_{k=1}^9\frac{1}{2}A_k r^{2k}+g\frac{8192 r^{14}+49152 r^{12}-135168r^{10}+184320 r^8-114752r^6+960 r^4+720 r^2}{\left(16 r^8-64 r^6+120 r^4+45\right)^2}. 
\end{equation}
The wavefunction is 
\begin{equation}\label{CE SUSY 4 wavefunction m1=2 m2=3 m3=4}
    \psi(r)=r^{l+1}{\cal W}_{(m_1,m_2,m_3,m_4)}^{\nu}(r)e^{\omega(r)}f(r),\quad \omega(r)=\sum_{p=1}^{5}{B}_p r^{2p},\quad \nu=\frac{1}{2}\bigg(1-\sqrt{1-4g}\bigg),
\end{equation}
the coefficients of the gauge factor are
\begin{equation}\label{CE SUSY 4 gauge factors coefficients}
\begin{split}
       B_5=-\frac{\sqrt{A_9}}{10},\quad B_4=-\frac{A_8}{16 \sqrt{A_9}}\quad  B_3=\frac{A_8^2-4 A_7 A_9}{48 A_9^{3/2}},\quad B_2=\frac{-A_8^3+4 A_7 A_9 A_8-8 A_6 A_9^2}{64 A_9^{5/2}},\\ B_1=\frac{5 A_8^4-24 A_7 A_9 A_8^2-64 A_5 A_9^3+16 \left(A_7^2+2 A_6 A_8\right) A_9^2}{256 A_9^{7/2}}. 
\end{split}
\end{equation}

\begin{figure}[H]
  \centering
  \begin{subfigure}{0.34\linewidth}
    \centering
    \includegraphics[width=\linewidth]{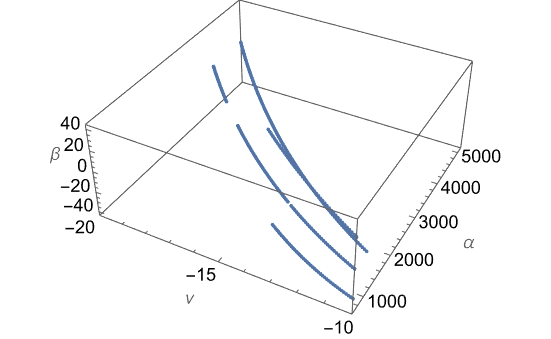}
    \caption{Allowed parameters for 2nd order SUSY deformation $(m_1=2,m_2=3)$}
    \label{f 2nd}
  \end{subfigure}
  \hfill
  \begin{subfigure}{0.32\linewidth}
    \centering
    \includegraphics[width=\linewidth]{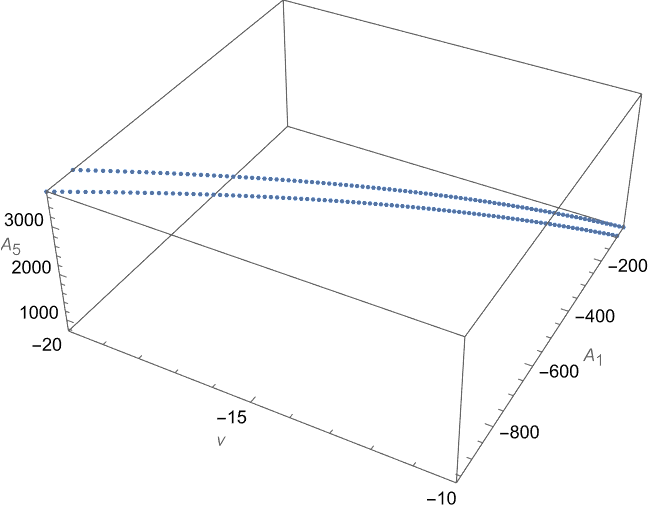}
    \caption{Allowed parameters for 3rd order SUSY deformation}
    \label{f 3rd}
  \end{subfigure}
    \hfill
  \begin{subfigure}{0.32\linewidth}
    \centering
    \includegraphics[width=\linewidth]{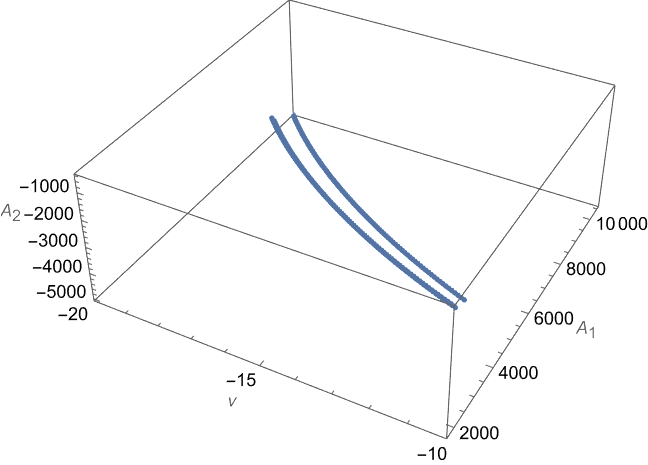}
    \caption{Allowed parameters for 4th order SUSY deforamtion }
    \label{f 4th}
  \end{subfigure}
  \caption{Allowed parameters in the QES potentials of the higher order SUSY transformation. }
  \label{fig higher order}
\end{figure}

Here we omit additional details on the three cases above, but provide evidence of their quasi-eact solvability through some numeric results for physical parameter in the parameter space and real energy spectrum.
Fig.(\ref{f 2nd}) provides parameter relations for the QES potential of 2nd order SUSY transformation in the Section \ref{higher order SUSY m1=2 m2=3} not belongs to the QES potential with $m_1=4,m_2=5$ above. We set $n=1, l=3$ and change $\nu$ from $-20$ to $-10$ for the $m_1=2,m_2=3$. The lines represent the solutions of the QES potential.  Fig.(\ref{f 3rd}) and Fig.(\ref{f 4th}) show the allow parameters for the 3rd order and 4th order SUSY transformation cases, respectively.

\bibliographystyle{myunsrt}
\bibliography{Reference}

\end{document}